\documentclass[12pt]{article}

\usepackage{amsfonts}
\usepackage{amssymb}
\usepackage{amsmath}

\newtheorem{definition}{Definition}
\newtheorem{lemma}{Lemma}
\newtheorem{theorem}{Theorem}
\newtheorem{proposition}{Proposition}
\newtheorem{corollary}{Corollary}
\newtheorem{property}{Property}

\begin{document}

\author{{Roustam Zalaletdinov} \\ [5mm]
\emph{Department of Mathematics and Statistics, Dalhousie University} \\
\emph{Chase Building, Halifax, Nova Scotia, Canada B3H 3J5} \\ [2mm]
\emph{Department of Theoretical Physics, Institute of Nuclear Physics} \\
\emph{Uzbek Academy of Sciences, Ulugbek, Tashkent 702132, Uzbekistan, CIS} \\
}
\title{{\LARGE \textbf{Space-time Averages of \\
Classical Physical Fields}}}
\date{}
\maketitle

\begin{abstract}
A review on the main results concerning the algebraic and differential
properties of the averaging and coordination operators and the properties of
the space-time averages of macroscopic gravity is given. The algebraic and
differential properties of the covariant space-time averaging procedure by
means of using the parallel transportation averaging bivector operator are
analyzed. The structure of the pseudo-Riemannian space-time manifolds of
general relativity averaged by means of this procedure is discussed. A
comparison of both procedures is given and the directions of further
development of space-time averaging procedures of the physical classical
fields are outlined.
\end{abstract}

\section{Introduction}

Space-time averaging procedures play an important role in modern physics
because of, first of all, their relevance to the derivation of classical
macroscopic theories. A well-known example of such a procedure is the
space-time scheme developed for averaging out the microscopic Lorentz
electrodynamics to derive the macroscopic Maxwell electrodynamics (see, for
example, \cite{Nova:1955}-\cite{Inga-Jami:1985})\footnote{
Alternative approaches for deriving macroscopic electrodynamics which apply
other averaging procedures (for example, space averaging and statistical
ensemble averaging) are not considered here. For such approaches and a
discussion about their interrelations, physical significance, etc. see \cite
{deGr:1969}-\cite{Rota:1960}, \cite{Russ:1970}-\cite{Jack:1975} and \cite
{Zala:1997}.}. Another important physical argument for considering
space-time averaging procedures is that they are relevant to modelling the
process of the physical space-time measurements. It is the space-time
averages of physical fields that are known \cite{Bohr-Rose:1933}, \cite
{DeWi:1962} to have direct observational status and physical meaning.

The space-time averaging procedure of classical electrodynamics utilizes
explicitly the flat character of the Minkowski space-time manifold and its
formulation is essentially based on the existence of Cartesian coordinates.
In this connection, the following questions are of interest and importance:
whether it is possible (\emph{a}) to encode the properties of the averaging
procedure in a covariant manner suitable for differentiable manifolds not
equipped with a metric and a connection in general, and (\emph{b}) to
formulate the corresponding covariant averaging procedure for the objects
which can live on such manifolds and to clear up its geometric meaning.
Having replied (\emph{a}) and (\emph{b}), the next question of primary
importance is (\emph{c}) how the averaging procedure may be made compatible
with a metric and a connection when the differentiable manifold possesses
those structures.

It should be pointed out here that the above problem of generalizing the
flat space-time procedure for curved manifolds goes far beyond being simply
an academic problem. One of its most important areas of applicability is the
general theory of relativity where the space-time is a 4-dimensional
pseudo-Riemannian manifold. As is well-known, there is not yet a
satisfactory derivation of the Maxwell equations in general relativity,
apart from the covariantization procedure where the partial derivatives of
the special theory of relativity are replaced by covariant ones (see, for
example, \cite{MTW:1973}). A physically motivated and mathematically correct
derivation of the macroscopic Maxwell equations by averaging out the general
relativistic microscopic Maxwell-Lorentz equations is still lacking and,
furthermore, the very foundations of the microscopic electrodynamics in
general relativity are not well-established yet. Another problem in general
relativity where the availability of a space-time averaging procedure is of
primary importance is the so-called averaging problem (see \cite{Zala:1997},
\cite{Elli:1984}-\cite{Kras:1996} for a review and discussion). Its main
motivation comes from cosmology where Einstein's equations are usually
utilized with a hydrodynamic stress-energy tensor without any satisfactory
proof of why the left-hand side of the equations (the field operator) keeps
the same structure while the right-hand side has been changed, or averaged,
from a real discrete matter distribution (stars, galaxies, etc.) to a
continuous one. The task here is to carry out a space-time averaging of
Einstein's equations in order to understand the structure of the averaged
(macroscopic) field equations and apply them to deal with the cosmological
problems which are especially related to the large-scale structure of our
Universe. A solution for this problem is also desirable in order to provide
a rigorous basis for constructing continuous matter models from discrete
ones. This kind of consideration is based in modern cosmology mainly on
phenomenological grounds.

The goal of this paper is threefold: (1) to give a review of the main
results concerning the algebraic structure of the averaging and coordination
operators and the properties of the space-time averages of macroscopic
gravity \cite{Zala:1997}, \cite{Zala:1992}, \cite{Zala:1993}-\cite
{Mars-Zala:1997}, (2) to analyze the algebraic and differential properties
of the covariant averaging procedure using the parallel transportation
averaging bivector operator and (3) to discuss the structure of the
pseudo-Riemannian space-time manifolds of general relativity averaged by
means of this procedure. The paper is organized as follows. Section 2 gives
an overview of the space-time averaging scheme used in the classical Maxwell
macroscopic electrodynamics. The space-time averages of macroscopic gravity
are defined in Section 3. The next Section is devoted to the algebra of the
averaging operator with a discussion of the algebraic properties of the
space-time averages. The differential properties of the averages and the
coordination bivector are summarized in Section 5. In Section 6 the
existence theorems for the averaging and coordination operators are given.
Section 7 is devoted to the definition and properties of the proper
coordinate systems, the coordinates in which the averaging operators take
their simplest form. The space-time covariant averaging procedure by means
of using the parallel transportation averaging bivector operator is defined
in Section 8 and its algebraic properties are discussed. The next Section
gives a proof of the uniqueness of these space-time averaging values. The
differential properties of the averages are studied in Section 10. In
Section 11 averaging out of the pseudo-Riemannian space-time manifolds of
general relativity is shown to lead to the non-Riemannian geometry of the
averaged space-times. The obtained results are discussed in Conclusions
where a comparison between the space-time averaging scheme of macroscopic
gravity and that based on the parallel transportation averaging bivector
operator is made and the directions of further development are outlined.

\section{The Space-time Averages in the Minkowski Space-time}

\label{*stamm}

The space-time averaging procedure applied in the classical Maxwell
macroscopic electrodynamics is explicitly based on the flat character of the
Minkowski space-time $\mathcal{E}$ and the existence of the Cartesian
coordinates $(t,x^{a}),$ $a=1,2,3$. The space-time averages are defined as
\cite{Nova:1955}-\cite{Inga-Jami:1985}, \cite{Mars-Zala:1997}.

\begin{definition}[The space-time averages in the Minkowski space-time]
The average value of a tensor field $p_{\beta }^{\alpha }(t,x^{a})$, $
(t,x^{a})$ $\in \mathcal{E}$, over a compact space region $S$ and a finite
time interval $T$ at a supporting point $(t,x^{a})\in T\times S$ is given by
\begin{equation}
\left\langle p_{\beta }^{\alpha }\right\rangle _{\mathcal{E}}(t,x^{a})=\frac{
1}{TV_{S}}\int_{T}\int_{S}p_{\beta }^{\alpha }(t+t^{\prime
},x^{a}+x^{a\prime })dt^{\prime }d^{3}x^{\prime }.  \label{defaver:ED}
\end{equation}
Here $V_{S}$ is the 3-volume of the region $S$, which is usually taken as a
3-sphere of radius $R$ around the point $x^{a}$ at the instant of time $t$,
\begin{equation*}
V_{S}=\int_{S}d^{3}x^{\prime }.
\end{equation*}
\end{definition}

Formula (\ref{defaver:ED}) defines the average value $\left\langle p_{\beta
}^{\alpha }\right\rangle _{\mathcal{E}}(t,x^{a})$ at a point $(t,x^{a})$ on
the manifold $\mathcal{E}$. In order to obtain an averaged tensor field $
\left\langle p_{\beta }^{\alpha }\right\rangle _{\mathcal{E}}(t,x^{a})$
defined at every point $(t,x^{a})\in \mathcal{E}$ and define its
derivatives, it is necessary to make additional assumptions concerning the
averaging regions $S$ and the intervals $T$. These assumptions are usually
made only tacitly (see, however, a discussion in \cite{Nova:1955}), or they
are supposed to be trivial, but it is necessary to write them down here
explicitly:

(\emph{i}) an interval $T$ and a region $S$ must be prescribed at every
point $(t,x^{a})$ $\in \mathcal{E}$ in order to define an averaged field $
\left\langle p_{\beta }^{\alpha }\right\rangle _{\mathcal{E}}(t,x^{a})$,

(\emph{ii}) all the regions $S$ and time intervals $T$ are typical in some
defined sense.

They are usually required to be of the same shape and volume, $V_{S}=\mathrm{
const}$ and $T=\mathrm{const}$, such as
\begin{equation}
\frac{\partial T}{\partial t}=0,\quad \frac{\partial T}{\partial x^{i}}
=0,\quad \frac{\partial V_{S}}{\partial t}=0,\quad \frac{\partial V_{S}}{
\partial x^{i}}=0,  \label{covering}
\end{equation}
at every point $(t,x^{a})$ $\in \mathcal{E}$ and related to each other by
shifting along the Cartesian coordinate lines.

These properties are very easily arranged by the Lie-dragging of an interval
$T$ and a region $S$, chosen around a point $(t,x^{a})$ $\in \mathcal{E}$,
along the congruences of the Cartesian coordinate lines\footnote{
A definition of Lie-dragging, or dragging, of a region of a manifold along a
vector field (congruence), or in other words, mapping of a region into
another along a vector field (congruence) can be found in any standard
textbook on differential geometry (see, for example, \cite{Schu:1980}, \cite
{Koba-Nomi:1963}). Throughout the paper Lie-draggings of regions are
supposed to be $C^{\infty }$-diffeomorphisms.} to obtain a \textquotedblleft
covering\textquotedblright\ of the manifold (or its connected part) with a
time interval of the same length and a region of the same shape and volume
around each point of $E$. Consequences of the properties (\emph{i}) and
(\emph{ii}) are, first of all, the uniqueness of the definition of the
average field $\left\langle p_{\beta }^{\alpha }\right\rangle _{\mathcal{E}
}(t,x^{a})$, and, secondly, the commutation formulae between the averaging
and the partial derivatives,
\begin{equation}
\frac{\partial }{\partial t}\left\langle p_{\beta }^{\alpha }\right\rangle _{
\mathcal{E}}(t,x^{a})=\left\langle \frac{\partial }{\partial t}p_{\beta
}^{\alpha }\right\rangle _{\mathcal{E}}(t,x^{a}),\quad \frac{\partial }{
\partial x^{a}}\left\langle p_{\beta }^{\alpha }\right\rangle _{\mathcal{E}
}(t,x^{a})=\left\langle \frac{\partial }{\partial x^{a}}p_{\beta }^{\alpha
}\right\rangle _{\mathcal{E}}(t,x^{a}).  \label{comm:ED}
\end{equation}
The essence of this approach lies in the exploitation of the calculational
advantages of the Cartesian coordinates on a flat manifold. The Cartesian
coordinates play a central role both in defining the averages (\ref
{defaver:ED}) and in obtaining properties, such as (\ref{comm:ED}), which
will allow the averaging out of the partial (ordinary) differential
equations under interest. It should be stressed that by requiring (\emph{i})
and (\emph{ii}) the averages $\left\langle p_{\beta }^{\alpha }\right\rangle
_{\mathcal{E}}(t,x^{a})$ become local functions of $(t,x^{a})$, i.e.
\begin{equation}
\left( \frac{\partial }{\partial x^{a}}\frac{\partial }{\partial t}-\frac{
\partial }{\partial t}\frac{\partial }{\partial x^{a}}\right) \left\langle
p_{\beta }^{\alpha }\right\rangle _{\mathcal{E}}(t,x^{a})=0,
\label{comm2:ED}
\end{equation}
and, therefore, they are provided with proper analytical properties.
Furthermore, the functional dependence of the average (\ref{defaver:ED}) on
the averaging region and interval, $S$ and $T$, becomes simply a parametric
dependence on the value of the volume $V_{S}$ and the length $T$\textrm{. }
One can, therefore, apply the standard differential and integral calculus to
deal with the averages and averaged equations and apply the methods of
mathematical physics to solve differential and integral equations for the
averages. The set of averaged components $\left\langle p_{\beta }^{\alpha
}\right\rangle _{\mathcal{E}}(t,x^{a})$ given by (\ref{defaver:ED}) in
Cartesian coordinates are the components of a Lorentz tensor within the
class of coordinate transformations $\widetilde{x}^{\alpha }=\Lambda _{\beta
}^{\alpha }x^{\beta }+a^{\beta }$ with a constant shift $a^{\beta }$ and a
constant Lorentz transformation matrix $\Lambda _{\beta }^{\alpha }$. Now,
if it is necessary to consider an average field in another coordinate
system, the components of $\left\langle p_{\beta }^{\alpha }\right\rangle _{
\mathcal{E}}(t,x^{a})$ in that new system are found by applying the standard
tensorial transformation law.

There is still another property of the averages $\left\langle p_{\beta
}^{\alpha }\right\rangle _{\mathcal{E}}(t,x^{a})$ widely used in the Maxwell
macroscopic electrodynamics (but again, often only tacitly assumed), namely,
the idempotency of the space-time averages
\begin{equation}
\left\langle \left\langle p_{\beta }^{\alpha }\right\rangle _{\mathcal{E}
}\right\rangle _{\mathcal{E}}(t,x^{a})=\left\langle p_{\beta }^{\alpha
}\right\rangle _{\mathcal{E}}(t,x^{a}).  \label{idem:ED}
\end{equation}
In order to prove this property we must calculate the average value $
\left\langle \left\langle p_{\beta }^{\alpha }\right\rangle _{\mathcal{E}
}\right\rangle _{\mathcal{E}}(t,x^{a})$ of the average $\left\langle
p_{\beta }^{\alpha }\right\rangle _{\mathcal{E}}(t,x^{a})$
\begin{eqnarray}
&&\left\langle \left\langle p_{\beta }^{\alpha }\right\rangle _{\mathcal{E}
}\right\rangle _{\mathcal{E}}(t,x^{a})=  \notag \\
&&\frac{1}{TV_{S}}\int_{T}\int_{S}\left( \frac{1}{T^{\prime }V_{S^{\prime }}}
\int_{T^{\prime }}\int_{S^{\prime }}p_{\beta }^{\alpha }(t+t^{\prime
}+t^{\prime \prime },x^{a}+x^{\prime a}+x^{\prime \prime a})dt^{\prime
\prime }d^{3}x^{\prime \prime }\right) dt^{\prime }d^{3}x^{\prime }.
\label{avav}
\end{eqnarray}
Now, the expression (\ref{avav}) leads to (\ref{idem:ED}) under either of
two additional assumptions: (\emph{iii}$^{\prime }$) the averaging region $
T^{\prime }\times S^{\prime }$ is the same region $T\times S$ with the
supporting point at $(t^{\prime },x^{\prime a})\in T\times S$ and the
average value $\left\langle p_{\beta }^{\alpha }\right\rangle _{\mathcal{E}
}(t,x^{a})$ does not depend on the choice of a supporting point within a
chosen region; (\emph{iii}$^{\prime \prime }$) the averaging region $
T^{\prime }\times S^{\prime }$ is a different neighboring region and the
average value $\left\langle p_{\beta }^{\alpha }\right\rangle _{\mathcal{E}
}(t,x^{a})$ remains the same if evaluated over any neighboring regions
containing a neighborhood of $x$. Although they seem different the above
assumptions are nevertheless essentially equivalent. Indeed, (\emph{iii} $
^{\prime }$) puts emphasis on the independence of the average value with
respect to the choice of a supporting point from a set of all possible
points in a fixed averaging region, while (\emph{iii}$^{\prime \prime }$)
puts emphasis on the independence of the average value with respect to the
choice of an averaging region from a set of all possible regions defined by
a neighborhood of a fixed supporting point. Remembering (\emph{ii}) that the
averaging regions are typical, change of the supporting point can be
considered as change of the averaging region, and vice versa. Both
assumptions thereby encode the same fundamental property of the averages
which is inherently related to the philosophy of averaging itself - an
averaging region is considered as a point from the macroscopic point of
view, and the change of either a supporting point or an averaging region
does not affect the corresponding average value. The reason for this is that
the differences caused by the above variations are negligibly small from a
macroscopic point of view.

It should be noted here that in classical hydrodynamics, as well as in the
macroscopic electrodynamics, a definition of an average (either over space,
time, ensemble, or a combination of such) and its properties are vital
elements of the theory itself for it is clearly understood that the form of
the equations depends on the definition and properties of the average. The
definition (\ref{defaver:ED}) under conditions (\emph{i}) and (\emph{ii})
with the properties (\ref{comm:ED}), (\ref{comm2:ED}) and (\ref{idem:ED}),
which are part of the Reynolds conditions in hydrodynamics, is known to
result in the Reynolds equations of hydrodynamics describing the dynamics of
turbulence. If one of the Reynolds conditions is absent one must get
different equations. For a discussion on averages and their properties in
hydrodynamics, see, for example, \cite{Moni-Yagl:1971} and references
therein.

Thus, the well-known procedure of space-time averaging in classical
electrodynamics presupposes the above specific conditions (\emph{i}),
(\emph{ii}) and (\emph{iii}) to ensure reasonable analytical and tensorial
properties of the averages (\ref{defaver:ED}). Clearly, the whole procedure
relies crucially on the existence and properties of the exceptional
coordinate system in a flat space-time, namely, Cartesian coordinates. These
conditions, which seem more or less trivial at first sight, require further
analysis to make clear their geometrical meaning and invariant content. This
will allow a reasonable formulation of a space-time averaging procedure on
general (not necessarily flat) manifolds.

\section{The Definition of Space-time Averages in Macroscopic Gravity}

\label{*sstamg}

Let us remind the definition of the space-time averages adopted in
macroscopic gravity \cite{Zala:1992}, \cite{Zala:1993}, \cite{Mars-Zala:1997}
. This procedure is a generalization of the space-time averaging procedure
adopted in electrodynamics (see Section \ref{*stamm}) and it is also based
on the concept of Lie-dragging of averaging regions, which makes it valid
for any differentiable manifold.

\begin{definition}[The space-time averages in macroscopic gravity]
Chosen a com-\linebreak pact region $\Sigma \subset \mathcal{M}$ in an $n$
-dimensional differentiable metric manifold $(\mathcal{M}$, $g_{\alpha \beta
})$ with a volume $n$-form, and a supporting point $x\in \Sigma $ to which
the average value will be prescribed, the space-time average value of an
object (tensor, geometric object, etc.) $p_{\beta }^{\alpha }(x),\,x\in
\mathcal{M\ }$, over a region $\Sigma $ at the supporting point $x\in \Sigma
$ is defined as
\begin{equation}
\left\langle p\right\rangle _{\beta }^{\alpha }(x)=\frac{1}{V_{\Sigma }}
\int_{\Sigma }\mathbf{p}_{\beta }^{\alpha }(x,x^{\prime })\sqrt{-g^{\prime }}
d^{n}x^{\prime }\equiv \langle \mathbf{p}_{\beta }^{\alpha }\rangle (x),
\label{defaver:MG}
\end{equation}
where $V_{\Sigma }$ is the volume of the region $\Sigma $,
\begin{equation}
V_{\Sigma }=\int_{\Sigma }\sqrt{-g}d^{n}x.  \label{volume}
\end{equation}
\end{definition}

Here the integration is carried out over all points $x^{\prime }\in \Sigma $
, $g^{\prime }=\det \left[ g_{\alpha \beta }(x^{\prime })\right] $\footnote{
Since the primary interest is in space-time manifolds, it is assumed that
this determinant is negative, otherwise the negative sign in $\sqrt{-g}$
must be dropped.} and the bold face object $\mathbf{p}_{\beta }^{\alpha
}(x,x^{\prime })$ in the integrand of (\ref{defaver:MG}) is a bilocal
extension of the object $p_{\beta }^{\alpha }(x)$ defined as
\begin{equation}
\mathbf{p}_{\beta }^{\alpha }(x,x^{\prime })=\mathcal{A}_{\mu ^{\prime
}}^{\alpha }(x,x^{\prime })p_{\nu ^{\prime }}^{\mu ^{\prime }}(x^{\prime })
\mathcal{A}_{\beta }^{\nu ^{\prime }}(x^{\prime },x),  \label{bilocext}
\end{equation}
by means of bilocal averaging operators $\mathcal{A}_{\beta ^{\prime
}}^{\alpha }(x,x^{\prime })$ and $\mathcal{A}_{\beta }^{\alpha ^{\prime
}}(x^{\prime },x)$. The averaging scheme is covariant and linear,
\begin{equation}
\left\langle p+q\right\rangle _{\beta }^{\alpha }(x)=\left\langle
p\right\rangle _{\beta }^{\alpha }+\left\langle q\right\rangle _{\beta
}^{\alpha }(x)\quad \mathrm{or}\quad \langle a\mathbf{p}_{\beta }^{\alpha }+b
\mathbf{q}_{\beta }^{\alpha }\rangle (x)=a\langle \mathbf{p}_{\beta
}^{\alpha }\rangle (x)+b\langle \mathbf{q}_{\beta }^{\alpha }\rangle (x),
\label{linear}
\end{equation}
where $a,b\in \mathrm{R}$, by construction and the averaged object $
\left\langle p\right\rangle _{\beta }^{\alpha }(x)$ keeps the same tensorial
character as $p_{\beta }^{\alpha }$.

Let us suppose that the bilocal functions $\mathcal{A}_{\beta ^{\prime
}}^\alpha (x,x^{\prime })$ and $\mathcal{A}_\beta ^{\alpha ^{\prime
}}(x^{\prime },x)$ are defined locally on an open subset $\mathcal{U}\subset
\mathcal{M}$, $x,x^{\prime }\in \mathcal{U}$. In the following sections
their algebraic and differential properties are formulated and analyzed to
show that averaging operators with such properties do exist and also to find
out the corresponding properties of the averages (\ref{defaver:MG}).

\section{The Algebra of the Bilocal Averaging Operator}

\label{*abao}

The following algebraic properties, which are a formalization of the
properties of the space-time averages in the Maxwell macroscopic
electrodynamics by applying the language of bilocal operators, are required
to hold \cite{Zala:1997}, \cite{Zala:1992}, \cite{Zala:1993},
\cite{Mars-Zala:1997}.

\begin{property}[The coincidence limit]
The coincidence limit of $\mathcal{A}_{\beta ^{\prime }}^{\alpha }$ is
\begin{equation}
\lim_{x^{\prime }\rightarrow x}\mathcal{A}_{\beta ^{\prime }}^{\alpha
}(x,x^{\prime })=\delta _{\beta }^{\alpha }.  \label{coincidence}
\end{equation}
\end{property}

\begin{property}[The idempotency]
The operator $\mathcal{A}_{\beta ^{\prime }}^{\alpha }$ is idempotent
\begin{equation}
\mathcal{A}_{\beta ^{\prime }}^{\alpha }(x,x^{\prime })\mathcal{A}_{\gamma
^{\prime \prime }}^{\beta ^{\prime }}(x^{\prime },x^{\prime \prime })=
\mathcal{A}_{\gamma ^{\prime \prime }}^{\alpha }(x,x^{\prime \prime }).
\label{idempotency}
\end{equation}
\end{property}

These two properties imply that $\mathcal{A}_{\beta }^{\alpha ^{\prime
}}(x^{\prime },x)$ is the inverse operator of $\mathcal{A}_{\beta ^{\prime
}}^{\alpha }(x,x^{\prime })$, $\mathcal{A}_{\beta ^{\prime }}^{\alpha }
\mathcal{A}_{\gamma }^{\beta ^{\prime }}=\delta _{\gamma }^{\alpha }$ and $
\mathcal{A}_{\beta ^{\prime }}^{\alpha }\mathcal{A}_{\alpha }^{\gamma
^{\prime }}=\delta _{\beta ^{\prime }}^{\gamma ^{\prime }}$, and that the
average tensor $\left\langle p_{\beta }^{\alpha }\right\rangle (x)$ takes
the same value as the original tensor $p_{\beta }^{\alpha }(x)$, $
\left\langle p_{\beta }^{\alpha }\right\rangle (x)=p_{\beta }^{\alpha }(x)$,
when the space-time averaging region $\Sigma $ is chosen infinitesimally
small, or it tends to zero. This implies an additional algebraic property of
the averages (\ref{defaver:MG}), namely, that the space-time averaging
procedure commutes with the operation of the index contraction
\begin{equation}
\left\langle p\right\rangle _{\alpha }^{\alpha }(x)=\langle \mathbf{p}
_{\alpha }^{\alpha }\rangle (x).  \label{contraction}
\end{equation}

The idempotency (\ref{idempotency}) of the averaging operator $\mathcal{A}
_{\beta ^{\prime }}^{\alpha }$ is designed to provide the idempotency of the
averages (\ref{defaver:MG}) in macroscopic gravity
\begin{equation}
\left\langle \left\langle p\right\rangle \right\rangle _{\beta }^{\alpha
}(x)=\left\langle p\right\rangle _{\beta }^{\alpha }(x),\quad \mathrm{or}
\quad \left\langle \langle \mathbf{p}_{\beta }^{\alpha }\rangle
\right\rangle (x)=\langle \mathbf{p}_{\beta }^{\alpha }\rangle (x).
\label{aver=aver}
\end{equation}
Indeed, let us consider for simplicity a vector $v^{\alpha }(x)$ and
calculate the twice averaged value $\left\langle \left\langle v^{\alpha
}\right\rangle \right\rangle (x)$ around the same point $x\in \mathcal{U}$
by using the definition (\ref{defaver:MG})
\begin{equation}
\left\langle \left\langle \mathbf{v}^{\alpha }\right\rangle \right\rangle
(x)=\frac{1}{V_{\Sigma _{x}}}\int_{\Sigma _{x}}\left( \frac{1}{V_{\Sigma
_{x^{\prime }}}}\int_{\Sigma _{x^{\prime }}}\mathcal{A}_{\beta ^{\prime
}}^{\alpha }(x,x^{\prime })\mathcal{A}_{\gamma ^{\prime \prime }}^{\beta
^{\prime }}(x^{\prime },x^{\prime \prime })v^{\gamma ^{\prime \prime
}}(x^{\prime \prime })\sqrt{-g^{\prime \prime }}d^{n}x^{\prime \prime
}\right) \sqrt{-g^{\prime }}d^{n}x^{\prime }  \label{avravr}
\end{equation}
where $\Sigma _{x^{\prime }}$ is an averaging region around the point $
x^{\prime }\in \mathcal{U}$. By applying the idempotency condition (\ref
{idempotency}) the expression (\ref{avravr}) takes the form
\begin{equation}
\left\langle \left\langle \mathbf{v}^{\alpha }\right\rangle \right\rangle
(x)=\frac{1}{V_{\Sigma _{x}}}\int_{\Sigma _{x}}\left( \frac{1}{V_{\Sigma
_{x^{\prime }}}}\int_{\Sigma _{x^{\prime }}}\mathcal{A}_{\gamma ^{\prime
\prime }}^{\alpha }(x,x^{\prime \prime })v^{\gamma ^{\prime \prime
}}(x^{\prime \prime })\sqrt{-g^{\prime \prime }}d^{n}x^{\prime \prime
}\right) \sqrt{-g^{\prime }}d^{n}x^{\prime }.  \label{vbarbar}
\end{equation}
Now, if the term in parentheses were independent of $x^{\prime }$, we could
take it outside the integral with respect to the variable $x^{\prime }$ to
get the resulting property (\ref{aver=aver}). This term in (\ref{vbarbar}),
however, depends explicitly on $x^{\prime }$ in the integration space-time
region $\Sigma _{x^{\prime }}$. Similarly as in the Maxwell macroscopic
electrodynamics, the idempotency (\ref{aver=aver}) of the averages (\ref
{defaver:MG}) follows under either of two additional assumptions \cite
{Mars-Zala:1997}, \cite{Zala:unpub}: (\emph{iii}$^{\prime }$) the averaging
region $\Sigma _{x^{\prime }}$ is the same region $\Sigma _{x}$ with the
supporting point at $x^{\prime }\in \Sigma _{x}$ and the average value $
\left\langle p\right\rangle _{\beta }^{\alpha }(x)$ does not depend on the
choice of a supporting point within a chosen region; (\emph{iii}$^{\prime
\prime }$) the averaging region $\Sigma _{x^{\prime }}$ is a different
neighboring region and the average value $\left\langle p\right\rangle
_{\beta }^{\alpha }(x)$ remains the same if evaluated over any neighboring
region containing a neighborhood of $x$. As it was emphasized in Section \ref
{*stamm} this is a fundamental property of any physically reasonable
classical averaging procedure and such a stability of averages comes from
the basic principles of averaging. Indeed, a microscopic field to be
averaged is supposed to have two essentially different variation scales
\footnote{
The case of two scales is discussed here for the sake of simplicity. Of
course, very often there is a hierarchy of scales, in which case the
arguments are applied for each couple of scales satisfying (\ref{region}) to
be micro- and macroscopic ones, respectively.}, $\lambda $ and $L$, such as $
\lambda <<L$ and an averaging region must be taken of an intermediate size $d
$ between the two scales,
\begin{equation}
\lambda <<d<<L,  \label{region}
\end{equation}
so that the averaging effectively smooths out all the variations of the
microscopic field of the scale $\lambda $. It is implicitly assumed in every
averaging scheme (for example, in the averaging schemes applied in classical
physics in Minkowski space-time, such as in hydrodynamics and
electrodynamics - see Section 2) that the result of the averaging is
insensitive to the choice of the supporting point within a fixed averaging
region and it is independent of the choice of integration (averaging) region
itself provided the scale $d$ satisfies the condition (\ref{region}). This
means that the microscopic averaging region is considered as a single
\textquotedblleft point\textquotedblright\ for the macroscopic averaged
field. Such regions have been called \textquotedblleft physically
infinitesimally small\textquotedblright\ by Lorentz \cite{Lore:1916}.

The trouble is, however, to perform a mathematically rigorous proof of this
fact. A satisfactory formal analysis of the idempotency property (\ref
{aver=aver}) of the space-time, time and space averages is still lacking,
though it is extensively used for the derivation of the averaged equations
in classical physics and its analysis\footnote{
It should be pointed out here that within the ensemble average procedure the
idempotency follows without problem. However, such averages have their own
problems. In particular, the whole body of problems related with idempotency
is replaced by the necessity to prove the ergodicity hypothesis which states
that ensemble and time (or space) averages are equivalent. Both ensemble and
volume averagings have their own advantages and areas of applicability in
describing physical phenomena. It is important to realize in this connection
that in all macroscopic settings a volume averaging (over space, time, or
space-time) is an unavoidable element \cite{deGr:1969}-\cite{Jack:1975}.}
(see, for example, \cite{Moni-Yagl:1971}, \cite{Lesl:1973}, \cite{Stan:1985}
in fluid mechanics and \cite{Nova:1955}, \cite{Inga-Jami:1985}, \cite
{deGr:1969}, \cite{Lore:1916} in the classical macroscopic electrodynamics).
It has been shown, however, that there are certain classes of volume
averaging procedures which satisfy the so-called Reynolds conditions (\ref
{comm:ED}), (\ref{linear}) and (\ref{aver=aver}), but these procedures and
corresponding averaging kernels do not have a clear physical interpretation
and they did not find direct applications in fluid mechanics (see \cite
{Kamp:1956}, \cite{Rota:1960} for a discussion and references).

As the consequences of requiring (\ref{coincidence}) and (\ref{idempotency})
in the structure of the bilocal operator is determined by the following
theorem \cite{Mars-Zala:1997}.

\begin{theorem}[The factorization of the averaging operator]
The bilocal operator\linebreak\ $\mathcal{A}_{\beta }^{\alpha ^{\prime
}}(x^{\prime },x)$ satisfying the Properties (\ref{coincidence}) and (\ref
{idempotency}) is idempotent if and only if it is factorized,
\begin{equation}
\mathcal{A}_{\beta }^{\alpha ^{\prime }}(x^{\prime },x)=F_{i}^{\alpha
^{\prime }}(x^{\prime })F_{\beta }^{-1i}(x),  \label{factor}
\end{equation}
where $F_{i}^{\beta }(x)$ is a set of $n$ linear independent vector fields
and $F_{\beta }^{-1i}(x)$ is the associated dual 1-form basis, $i=1,...,n$,
and summation over $i$ is carried out.
\end{theorem}

\section{The Differential Properties of the Bilocal Operators}

\label{*dpbo}

In this section the differential properties of the bilocal operators are
summarized. This overview is essential for understanding further results
concerning the space-time averaging scheme.

In order to obtain the averaged fields of the geometric objects on $\mathcal{
M}$ one needs to assign an averaging region $\Sigma _{x}$ to each point $x$
of $\mathcal{U}\subset \mathcal{M}$, where the averaging integral (\ref
{defaver:MG}) is to be evaluated. Furthermore, to calculate directional,
partial and covariant derivatives of the averaged fields a law of
correspondence between neighboring averaging regions must be defined \cite
{Zala:1992}, \cite{Zala:1993}, \cite{Mars-Zala:1997}, \cite{AZS:1990b}. A
natural way to define such a correspondence is to relate averaging regions
by Lie-dragging, or mapping of a region into another along a vector field,
see Section \ref{*stamm}, by means of another bilocal operator $\mathcal{W}
_{\beta }^{\alpha ^{\prime }}(x^{\prime },x)$ which is also assumed to
satisfy the coincidence limit property (\ref{coincidence}).

To derive the commutation formulae between the averaging and the derivation,
one should define first the directional derivative of an average field $
\left\langle p\right\rangle _{\beta }^{\alpha }(x)$ along a vector field $
\vec{\xi}=d/d\lambda $,
\begin{equation}
\frac{d}{d\lambda }\left\langle p\right\rangle _{\beta }^{\alpha
}(x)=\lim_{\Delta \lambda \rightarrow 0}\frac{1}{\Delta \lambda }\left[
\left\langle p\right\rangle _{\beta }^{\alpha }(x+\Delta x)-\left\langle
p\right\rangle _{\beta }^{\alpha }(x)\right] ,  \label{derivative}
\end{equation}
where $\Delta x^{\alpha }=\xi ^{\alpha }(x)\Delta \lambda $. Let us define
now the shift field for every point $x^{\prime }\in \Sigma _{x}$ as \cite
{Zala:1992}, \cite{Zala:1993}, \cite{Mars-Zala:1997}, \cite{AZS:1990b},
\begin{equation}
S^{\alpha ^{\prime }}(x^{\prime },x)=\mathcal{W}_{\beta }^{\alpha ^{\prime
}}(x^{\prime },x)\xi ^{\beta }(x).  \label{shift}
\end{equation}
Now, the averaging region $\Sigma _{x+\Delta x}$ associated with the point $
x+\Delta x$ is obtained by Lie-dragging the averaging region $\Sigma _{x}$ a
parametric length $\Delta \lambda $ (the same for all $x^{\prime }\in \Sigma
_{x}$) along the integral lines of the field $S^{\alpha ^{\prime }}$ by
means of a bilocal operator $\mathcal{W}_{\beta }^{\alpha ^{\prime
}}(x^{\prime },x)$.

The coordination bivector $\mathcal{W}_{\beta }^{\alpha ^{\prime }}$ allows
the construction of the shift vector $S^{\alpha ^{\prime }}$ for any
averaging region and any vector $\xi ^{\alpha }$, the shift vector at the
supporting point $x$ for that region, $x\in \Sigma _{x}$. By choosing $n$
such linearly independent vector fields $\xi _{i}^{\alpha }$ and shifting
averaging regions along them one can build a covering of the manifold with
one averaging region associated to every $x\in \mathcal{U}\subset \mathcal{M}
$. This procedure is a formalization of the condition (\emph{i}) of Section
\ref{*stamm}.

As a consequence of the definitions (\ref{defaver:MG}), (\ref{derivative})
and (\ref{shift}) one can obtain the following formula for the commutation
of partial differentiation and averaging \cite{Zala:1992}, \cite{Zala:1993},
\cite{Mars-Zala:1997}:
\begin{equation}
\left\langle p\right\rangle _{\beta ,\lambda }^{\alpha }=\langle \mathcal{A}
_{\mu ^{\prime }}^{\alpha }p_{\nu ^{\prime },\epsilon ^{\prime }}^{\mu
^{\prime }}\mathcal{A}_{\beta }^{\nu ^{\prime }}\mathcal{W}_{\lambda
}^{\epsilon ^{\prime }}\rangle +\langle \mathbf{p}_{\beta }^{\alpha }
\mathcal{W}_{\lambda :\epsilon ^{\prime }}^{\epsilon ^{\prime }}\rangle
-\left\langle p\right\rangle _{\beta }^{\alpha }\langle \mathcal{W}_{\lambda
:\epsilon ^{\prime }}^{\epsilon ^{\prime }}\rangle -\langle \mathcal{S}
^{\alpha }{}_{\sigma \lambda }\mathbf{p}_{\beta }^{\sigma }\rangle +\langle
\mathbf{p}_{\sigma }^{\alpha }\mathcal{S}^{\sigma }{}_{\beta \lambda
}\rangle .  \label{derivative2}
\end{equation}
Here, $\mathcal{S}^{\alpha }{}_{\sigma \lambda }$ are the so-called
structural functions,
\begin{equation}
\mathcal{S}^{\alpha }{}_{\beta \gamma }=\mathcal{A}_{\epsilon ^{\prime
}}^{\alpha }(\mathcal{A}_{\beta ,\gamma }^{\epsilon ^{\prime }}+\mathcal{A}
_{\beta ,\sigma ^{\prime }}^{\epsilon ^{\prime }}\mathcal{W}_{\gamma
}^{\sigma ^{\prime }}),  \label{structure-func}
\end{equation}
and $\mathcal{W}_{\lambda :\epsilon ^{\prime }}^{\epsilon ^{\prime }}$ is
the divergence of the coordination bivector, $\mathcal{W}_{\lambda :\epsilon
^{\prime }}^{\epsilon ^{\prime }}=\mathcal{W}_{\lambda ,\epsilon ^{\prime
}}^{\epsilon ^{\prime }}+(\ln \sqrt{-g})_{,\epsilon ^{\prime }}\mathcal{W}
_{\lambda }^{\epsilon ^{\prime }}$. In an affine connection space with
connection coefficients $\Gamma _{\beta \gamma }^{\alpha }$ (a Riemannian
space is considered in \cite{AZS:1990b}) the same formula (\ref{derivative2}
) remains valid for the covariant derivatives, where the partial
differentiation is replaced by the covariant one and the divergence $
\mathcal{W}_{\lambda :\epsilon ^{\prime }}^{\epsilon ^{\prime }}$ is
replaced by $\mathcal{W}_{\lambda ;\epsilon ^{\prime }}^{\epsilon ^{\prime
}}=\mathcal{W}_{\lambda ,\epsilon ^{\prime }}^{\epsilon ^{\prime }}+\Gamma
_{\epsilon ^{\prime }}\mathcal{W}_{\lambda }^{\epsilon ^{\prime }}$ (with $
\Gamma _{\epsilon ^{\prime }}=\Gamma _{\epsilon ^{\prime }\alpha ^{\prime
}}^{\alpha ^{\prime }}$). The commutation formula (\ref{derivative2}) has a
very transparent meaning: the first term in the right-hand side is the
average value of the derivative of $p_{\beta }^{\alpha }$ weighted by $
\mathcal{W}_{\beta }^{\alpha ^{\prime }}$, the last two terms are due to the
nontrivial averaging operator $\mathcal{A}_{\beta }^{\alpha ^{\prime }}$,
while the second and third terms describe the effect of a nontrivial
averaging measure in (\ref{defaver:MG}) and the variation in the value of
the volume for different regions, respectively. Indeed, the change in the
volume $V_{\Sigma }$ of an (averaging) region $\Sigma $ $\subset \mathcal{M}$
Lie-dragged along a vector field $\xi $ is given by, see, for example, \cite
{Schu:1980}, \cite{AZS:1990a},
\begin{equation}
\frac{d}{d\lambda }V_{\Sigma }=\int_{\Sigma }\mathrm{div}\,\xi d\Omega ,
\label{volchange}
\end{equation}
which can be written in terms of partial derivatives after using (\ref{shift}
) as
\begin{equation}
V_{\Sigma ,\beta }=\langle \mathcal{W}_{\beta :\alpha ^{\prime }}^{\alpha
^{\prime }}\rangle V_{\Sigma }\,.  \label{shift:volume}
\end{equation}

The expression (\ref{derivative2}) is the most general version of formulae
(\ref{comm:ED}) for arbitrary averaging and coordination operators.

A fundamental problem one has to face now is whether it is possible or not
to define a unique covering of the manifold $\mathcal{M}$, like in Minkowski
space-time (see Section \ref{*stamm}). In other words, given a microscopic
tensor field $p_{\beta }^{\alpha }(x)$, the goal is to determine a uniquely
defined averaged tensor field $\left\langle p\right\rangle _{\beta }^{\alpha
}(x)$ with reasonable analytical properties in its dependence on the
supporting point. Furthermore, we must deal with the problem that the
averages (\ref{defaver:MG}) depend functionally on the averaging regions $
\left\langle p\right\rangle _{\beta }^{\alpha }(x)=\left\langle
p\right\rangle _{\beta }^{\alpha }(x)[\Sigma ]$ and applying the commutation
formula (\ref{derivative2}) to find averaged equations would, in general,
bring volume dependent terms into them, which is undesirable. In the case of
Minkowski space-time the uniqueness of the averaged field with the proper
analytical behavior (\ref{comm2:ED}) and the parametric dependence on the
averaging region volume are ensured by the condition (\emph{ii}) (see
Section 2) which sets the averaging regions to be typical, of the same shape
and volume, $V_{S}=\mathrm{const}$ and $T=\mathrm{const}$, and shifted along
the Cartesian coordinate lines. To arrange similar properties for the
generalized averages (\ref{defaver:MG}) one must look for some specific
conditions on the coordination operator $\mathcal{W}_{\beta }^{\alpha
^{\prime }}$. The following remarkable theorem holds \cite{Zala:1992}, \cite
{Zala:1993}, \cite{Mars-Zala:1997} (its version for Riemannian manifolds has
been given in \cite{AZS:1990b}).

\begin{theorem}[The locality of the space-time average fields]
In the averaging region coordination by the bivector $\mathcal{W}_{\beta
}^{\alpha ^{\prime }}$ on an arbitrary differentiable manifold, it is
necessary and sufficient to require
\begin{equation}
\mathcal{W}_{[\beta ,\gamma ]}^{\alpha ^{\prime }}+\mathcal{W}_{[\beta
,\delta ^{\prime }}^{\alpha ^{\prime }}\mathcal{A}_{\gamma ]}^{\delta
^{\prime }}=0,  \label{diffW:1}
\end{equation}
for the average tensor field $\left\langle p\right\rangle _{\beta }^{\alpha
}(x)$ to be a single valued local function of the supporting point $x$ on $
\mathcal{U}\subset \mathcal{M}$,
\begin{equation}
\left\langle p\right\rangle _{\beta ,[\mu \nu ]}^{\alpha }=0.
\label{averlocal}
\end{equation}
\end{theorem}

Geometrically, (\ref{diffW:1}) means that, given an averaging region $\Sigma
$, the region $\widetilde{\Sigma }$ obtained by transporting $\Sigma $ along
an infinitesimal parallelogram constructed from two commuting vector fields $
\xi $ and $\zeta $ according to the law (\ref{shift}) coincides with the
original region, $\widetilde{\Sigma }=\Sigma $. This is a highly non-trivial
property which allows to construct a covering of the manifold with an
averaging region attached to every point in the manifold, thus generalizing
the corresponding part of the condition (\emph{ii}) in Section \ref{*stamm}.
In the formalism of bilocal exterior calculus the condition (\ref{diffW:1})
reads that the operator $\mathcal{W}_{\beta }^{\alpha ^{\prime }}$ is
biholonomic, which means that the bilocal coordinate 1-form basis $\mathcal{W
}^{\alpha ^{\prime }}=\mathcal{W}_{\beta }^{\alpha ^{\prime }}\mathrm{d}
x^{\beta }$ has vanishing biholonomicity coefficients in the bilocal
Maurer-Cartan equations \cite{Zala:1993}.

It should be noted here that formula (\ref{averlocal}) is analogous to
formula (\ref{comm2:ED}) in both form and meaning.

Another condition on the coordination bivector $\mathcal{W}_{\beta }^{\alpha
^{\prime }}$ is the requirement that the Lie-dragging of a region is a
volume-preserving diffeomorphism \cite{Zala:1992}, \cite{Zala:1993} (\cite
{AZS:1990b} for Riemannian manifolds)
\begin{equation}
\mathcal{W}_{\beta :\alpha ^{\prime }}^{\alpha ^{\prime }}=0,
\label{diffW:2}
\end{equation}
which means that the averaging regions do not change the value of the volume
when shifted (coordinated) along a chosen vector field $\xi $ according to
(\ref{shift}). This generalizes the corresponding part of the condition
(\emph{ii}) in Section 2.

Thus, the condition (\ref{diffW:1}) states that the average tensor field is
a single valued local function of the supporting point $x$ (\ref{averlocal}
). Adding the condition (\ref{diffW:2}), the average tensor field does not
depend explicitly on the value of the region volume $V=V_{\Sigma }$, and $V$
itself is a free parameter of the theory. Given a microscopic tensor field $
p_{\beta }^{\alpha }(x)$ on $\mathcal{M}$, the average tensor field $
\left\langle p\right\rangle _{\beta }^{\alpha }(x)$ is therefore uniquely
defined on $\mathcal{U}\subset \mathcal{M}$ and can be handled within the
framework of standard differential and integral calculus. Requiring
additionally that the two bivectors $\mathcal{A}_{\beta }^{\alpha ^{\prime
}} $ and $\mathcal{W}_{\beta }^{\alpha ^{\prime }}$ coincide,
\begin{equation}
\mathcal{A}_{\beta }^{\alpha ^{\prime }}=\mathcal{W}_{\beta }^{\alpha
^{\prime }},  \label{AequalW}
\end{equation}
the first term in the commutation formula (\ref{derivative2}) becomes
exactly the average derivative. Using all conditions (\ref{diffW:1}), (\ref
{diffW:2}) and (\ref{AequalW}), the commutation formula acquires a
remarkable simply form \cite{Zala:1992}, \cite{Zala:1993}, \cite
{Mars-Zala:1997},
\begin{equation}
\left\langle p\right\rangle _{\beta ,\gamma }^{\alpha }=\langle \mathbf{p}
_{\beta ,\gamma }^{\alpha }+\mathbf{p}_{\beta ,\alpha ^{\prime }}^{\alpha }
\mathcal{W}_{\gamma }^{\alpha ^{\prime }}\rangle .\,  \label{comm}
\end{equation}
The corresponding analogues of this expression for covariant differentiation
are obtained by replacing partial derivatives by covariant ones. To obtain
the expression for directional derivatives we must contract this expression
with a vector $\xi ^{\gamma }$ and insert the vector field $S^{\alpha
^{\prime }}$ from (\ref{shift}) in the second term of the right-hand side of
(\ref{comm}). Formula (\ref{comm}) generalizes formulae (\ref{comm:ED}) and
it can be easily shown \cite{Zala:1993} to become exactly (\ref{comm:ED}) if
$\mathcal{W}_{\beta }^{\alpha ^{\prime }}=\delta _{\beta }^{\alpha }$ and
the volume $n$-form $\varepsilon $ is standard, that is $(\ln \sqrt{-g}
)_{,\epsilon ^{\prime }}=0$ and $\varepsilon =\mathrm{d}x^{1}\wedge
...\wedge \mathrm{d}x^{n}$ \cite{Schu:1980}, \cite{Koba-Nomi:1963}. In
Section VIII one can find more details on this particular case, see formulae
(\ref{defaver:MG:prop}) and (\ref{commu:2}).

\section{The Existence Theorems}

\label{*et}

The differential conditions (\ref{diffW:1}) and (\ref{diffW:2}) together
with the algebraic conditions (\ref{coincidence}), (\ref{idempotency}) and
(\ref{AequalW}) are to be considered as a set of partial differential and
algebraic equations for the unknown functions $\mathcal{W}_{\beta }^{\alpha
^{\prime }}$. Provided a solution for the system is found, the existence of
such operators $\mathcal{A}_{\beta }^{\alpha ^{\prime }}$ and $\mathcal{W}
_{\beta }^{\alpha ^{\prime }}$ (and therefore of the averages with the above
described properties) is proved. Theorem (\ref{factor}) has revealed the
structure of the operator $\mathcal{W}_{\beta }^{\alpha ^{\prime }}$ obeying
the algebraic properties (\ref{coincidence}), (\ref{idempotency}). The
following theorem \cite{Mars-Zala:1997} gives the general solution of (\ref
{diffW:1}) (in \cite{Zala:1992}, \cite{Zala:1993} and \cite{AZS:1990b} for
Riemannian manifolds, the same theorem has proved a solution of (\ref
{diffW:1}) with (\ref{AequalW}) sought in a factorized form (\ref{factor}),
and now, with Theorem (\ref{factor}) taken into account, it gives the
general solution).

\begin{theorem}[The existence of space-time averaging operators]
In an arbitrary $n$-dimensional differentiable manifold the general solution
of the equations
\begin{equation}
\mathcal{W}_{[\beta ,\gamma ]}^{\alpha ^{\prime }}+\mathcal{W}_{[\beta
,\delta ^{\prime }}^{\alpha ^{\prime }}\mathcal{W}_{\gamma ]}^{\delta
^{\prime }}=0,  \label{diffWW:1}
\end{equation}
for the idempotent bilocal operator $\mathcal{W}_{\beta }^{\alpha ^{\prime
}}(x^{\prime },x)$ is given by
\begin{equation}
\mathcal{W}_{\beta }^{\alpha ^{\prime }}(x^{\prime },x)=f_{i}^{\alpha
^{\prime }}(x^{\prime })f_{\beta }^{-1i}(x)  \label{W}
\end{equation}
where $f_{i}^{\alpha }(x)\partial _{\alpha }=\mathbf{f}_{i}$ is any vector
basis satisfying the commutation relations,
\begin{equation}
\left[ \mathbf{f}_{i},\mathbf{f}_{j}\right] =C_{ij}^{k}\mathbf{f}_{k},
\label{commff}
\end{equation}
with constant structure functions (anholonomicity coefficients) $C_{ij}^{k}$
,
\begin{equation}
C_{ij}^{k}=\mathrm{const}.  \label{C:const}
\end{equation}
\end{theorem}

The next theorem proves the existence of solutions for the equation (\ref
{diffW:2}) within the class of bivectors satisfying (\ref{W}) \cite
{Mars-Zala:1997}, \cite{Zala:unpub} (its version for a particular subclass
of (\ref{W}), see Section \ref{*psc}, has been given in \cite{Zala:1992},
\cite{Zala:1993} and \cite{AZS:1990b} for the case of Riemannian manifolds).

\begin{theorem}[The existence of volume-preserving bivectors]
In an arbitrary $n$-di-\linebreak\ mensional differentiable metric manifold $
(\mathcal{M}$, $g_{\alpha \beta })$ with a volume $n$-form there always
exist locally volume-preserving bivectors $\mathcal{W}_{\beta }^{\alpha
^{\prime }}(x^{\prime },x)$ of the form (\ref{W}) with (\ref{C:const})
satisfying (\ref{diffW:2}).
\end{theorem}

It should be stressed here that the conditions $C_{jk}^{i}=\mathrm{const}$
due to Theorem (\ref{diffWW:1}) are essential for the proof of the last
Theorem and they guarantee the local existence of $n$ linear independent
divergence free vectors, the result holding for both orientable and
nonorientable manifolds (by using the so-called odd volume $n$-form \cite
{deRh:1960}, \cite{Mose:1965})\footnote{$n$ linear independent divergence
free vector fields on an $n$-dimensional differentiable manifold with a
volume $n$-form with structure functions (\ref{commff}) can be shown to
exist locally iff $C_{lk,i}^{i}=0$. This condition is fulfilled also
globally on parallelizable manifolds (both orientable and non-orientable)
\cite{Grom:1986}. This shows that Theorem of the existence of
volume-preserving bivectors is valid globally for such manifolds (a manifold
is called parallelizable if its tangent bundle is trivial).}.

Another important point is that $n$ vector fields $f_{i}^{\alpha }$
satisfying (\ref{commff}) and (\ref{C:const}) define a finite dimensional
Lie group on the averaged manifolds which is directly related with the
symmetries of such manifolds.

These two theorems prove the existence of solutions for the set of equations
(\ref{coincidence}), (\ref{idempotency}), (\ref{diffW:1}), (\ref{diffW:2}),
and (\ref{AequalW}), and therefore prove the existence of the bilocal
operator $\mathcal{W}_\beta ^{\alpha ^{\prime }}(x^{\prime },x)$ with the
corresponding algebraic and differential properties (\ref{averlocal}) and (
\ref{comm}) for the averages (\ref{defaver:MG}).

Now a particular subclass of the operators (\ref{W}) with (\ref{C:const})
will be considered to analyze some additional properties and to reveal the
functional structure of the subclass.

\section{The Proper Systems of Coordinates}

\label{*psc}

As it has been emphasized in the Introduction and Section 2, the space-time
averaging procedure adopted in electrodynamics is essentially formulated in
Cartesian coordinates and all its properties are shown by exploiting the
exceptional character of this coordinates (see Section II for details). The
covariant formalism developed for the averages (\ref{defaver:MG}) in
macroscopic gravity, see Sections 3-6, generalizes the averaging scheme of
macroscopic electrodynamics for arbitrary $n$-dimensional differentiable
manifolds and while keeping covariant properties which are analogous to
those in electrodynamics.

Let us now consider the macroscopic gravity averaging scheme for a
particular subclass of operators (\ref{W}) with (\ref{C:const}). This
particular subclass admits a special coordinate system in which the averages
and their properties have especially simple form and meaning. Such a
coordinate system is an analogue for macroscopic gravity of the Cartesian
coordinates in Minkowski space-time.

Let us hereby restrict the class of solutions of the equations (\ref
{diffWW:1}) to the subclass satisfying
\begin{equation}
\left[ \mathbf{f}_{i},\mathbf{f}_{j}\right] =0,  \label{coorff}
\end{equation}
that is $C_{ij}^{k}\equiv 0$. In this case the vector fields $f_{i}^{\alpha }
$ constitute a coordinate system and there always exist $n$ functionally
independent scalar functions $\phi ^{i}(x)$ such that the vector and
corresponding dual 1-form bases are of the form
\begin{equation}
f_{i}^{\alpha }(x(\phi ^{k}))=\frac{\partial x^{\alpha }}{\partial \phi ^{i}}
,\quad f_{\beta }^{-1i}(\phi (x^{\mu }))=\frac{\partial \phi ^{i}}{\partial
x^{\alpha }}.  \label{bases:coor}
\end{equation}
Thus, the bilocal operator $\mathcal{W}_{\beta }^{\alpha ^{\prime
}}(x^{\prime },x)$ becomes
\begin{equation}
\mathcal{W}_{\beta }^{\alpha ^{\prime }}(x^{\prime },x)=\frac{\partial
x^{\alpha ^{\prime }}}{\partial \phi ^{i}}\frac{\partial \phi ^{i}}{\partial
x^{\beta }}.  \label{W:coor}
\end{equation}
Being functionally independent, the set of $n$ functions $\phi ^{i}(x)$ can
be taken as a system of local coordinates on the manifold $\mathcal{M}$ \cite
{Zala:1992}, \cite{Zala:1993} (\cite{AZS:1990b} for Riemannian manifolds).

\begin{definition}[The proper coordinates]
A coordinate system $\{\phi ^{i}\}$ defined by $n$ scalar functions $\phi
^{i}=\phi ^{i}(x)$ in (\ref{bases:coor}) will be called a proper coordinate
system.
\end{definition}

The usefulness of this definition is motivated by the fact that in a proper
coordinate system the bilocal operator $\mathcal{W}_{\beta }^{\alpha
^{\prime }}(x^{\prime },x)$ takes the simplest possible form
\begin{equation}
\mathcal{W}_{j}^{i}(\phi ^{\prime },\phi )\equiv \mathcal{W}_{\beta
}^{\alpha ^{\prime }}(x^{\prime },x)_{\mid x^{\alpha }=\phi ^{i}}=\mathbf{
\delta }_{\beta }^{\alpha ^{\prime }}\equiv \mathbf{\delta }_{j}^{i},
\label{W:delta}
\end{equation}
where the bilocal Kronecker symbol $\mathbf{\delta }_{\beta }^{\alpha
^{\prime }}$ is defined as $\mathbb{\delta }_{\beta }^{\alpha ^{\prime
}}=\delta _{i}^{\alpha ^{\prime }}\delta _{\beta }^{i}$. The definition of
the average (\ref{defaver:MG}) acquires a remarkable simple form (closely
resembling the space-time averages of macroscopic electrodynamics (\ref
{defaver:ED})) when written using a proper coordinate system
\begin{equation}
\left\langle p\right\rangle _{j}^{i}(\phi )=\frac{1}{V_{\Sigma _{\phi }}}
\int_{\Sigma _{\phi }}p_{j}^{i}(\phi ^{\prime })\sqrt{-g(\phi ^{\prime })}
d^{n}\phi ^{\prime }.  \label{defaver:MG:prop}
\end{equation}

The Theorem of Section \ref{*et} has proved the existence of
volume-preserving bilocal operators (\ref{W}) and (\ref{C:const}), and
therefore the existence of solutions of (\ref{diffW:2}) for operators $
\mathcal{W}_{\beta }^{\alpha ^{\prime }}$ of the form (\ref{W:coor}) from
this theorem. It is useful, however, to prove it here independently due to
the above mentioned importance of the proper system of coordinates for the
macroscopic gravity averaging scheme. The following statement holds in this
case \cite{Zala:1992}, \cite{Zala:1993}, \cite{Mars-Zala:1997} (\cite
{AZS:1990b} for Riemannian manifolds).

\begin{proposition}[The existence of the proper coordinates]
In an arbitrary $n$-dime-\linebreak\ nsional differentiable metric manifold $
(\mathcal{M}$, $g_{\alpha \beta })$ with a volume $n$-form, there always
exist locally a set of $n$ scalar functionally independent functions $\phi
^{i}(x)$ such that the corresponding coordination bivector (\ref{W:coor})
satisfies condition (\ref{diffW:2}).
\end{proposition}

The following result is an obvious consequence of this Proposition.

\begin{corollary}[The volume-preserving proper coordinates]
Any proper coordinate system such that the corresponding bivector (\ref
{W:coor}) satisfies condition (\ref{diffW:2}) is a volume-preserving system
of coordinates,
\begin{equation}
\left( \mathrm{ln}\sqrt{-g(\phi ^{k})}\right) _{,j}=0\ ,\quad \mathrm{or}
\quad g(\phi ^{k})=\mathrm{const}.  \label{vol-pres}
\end{equation}
\end{corollary}

It should be noted here that all arguments concerning nonorientable
manifolds and global existence given after the Theorem of the existence of
volume-preserving bivectors apply here as well, see Section \ref{*et}.

For the case of (pseudo)-Riemannian manifolds, this Corollary states that in
a proper coordinate system the Christoffel symbols $\Gamma _{\beta \alpha
}^{\alpha }$, (which are $\Gamma _{\beta \alpha }^{\alpha }=(\mathrm{ln}
\sqrt{-g})_{,\beta }$ due to equi-affinity of Riemannian manifolds) vanish

\begin{equation}
\Gamma _{i}(\phi )\equiv \Gamma _{ij}^{j}=\left( \mathrm{ln}\sqrt{-g(\phi
^{k})}\right) _{,j}=0.  \label{vol-pres:Riem}
\end{equation}

Another useful characterization of the volume-preserving coordinates, in
addition to (\ref{vol-pres}), or (\ref{vol-pres:Riem}), can be obtained in
terms of the expansion of the vector fields tangent to the coordinates
lines. Defining a vector $\mathbf{\chi }_{(i)}$ tangent to a coordinate line
$\phi ^{i}$ as
\begin{equation}
\mathbf{\chi }_{(i)}=\chi _{(i)}^{j}\frac{\partial }{\partial \phi ^{j}}
=\delta _{(i)}^{j}\frac{\partial }{\partial \phi ^{j}},  \label{tangent}
\end{equation}
it is immediate to find that in the proper coordinate system $\{\phi ^{i}\}$
the expansion $\mathrm{div}{\chi }_{(i)}$ of the vector field (\ref{tangent}
) is
\begin{equation}
{\chi }_{(i):j}^{j}=\left( \mathrm{ln}\sqrt{-g(\phi ^{k})}\right) _{,i}=0,
\end{equation}
so that the condition of vanishing expansion for the tangent vector fields $
\mathbf{\chi }_{(i)}$ is equivalent to the definition of volume-preserving
coordinates (\ref{vol-pres}).

In addition to very simple and transparent forms of the coordination
bivector (\ref{W:coor}) and the averages (\ref{defaver:MG:prop}), the
volume-preserving proper coordinate systems also allow a remarkably simple
expression for the commutation between partial differentiation and
averaging. Indeed, expression (\ref{comm}) becomes in the proper coordinate
system
\begin{equation}
\frac{\partial }{\partial \phi ^{k}}\left\langle p\right\rangle
_{j}^{i}(\phi )=\left\langle \frac{\partial }{\partial \phi ^{\prime k}}
p_{j}^{i}(\phi ^{\prime })\right\rangle ,  \label{commu:2}
\end{equation}
which is exactly the same commutation formula for as in the averaging scheme
in Minkowski manifolds (\ref{comm:ED}).

Let us now study the functional structure of the class of the
volume-preserving coordinates to understand how large it is and how much
freedom for coordinate transformations it contains. The following
Proposition reveals the structure of the class \cite{Mars-Zala:1997}.

\begin{proposition}[The freedom of the proper coordinates]
The class of volume-pre-\linebreak\ serving coordinate transformations on an
arbitrary $n$-dimensional differentiable metric manifold $(\mathcal{M}$, $
g_{\alpha \beta })$ with a volume $n$-form, contains $(n-1)$ arbitrary
functions of $n$ arguments and one arbitrary function of $(n-1)$ arguments.
\end{proposition}

The set of proper coordinate systems forms quite a big class and this
functional freedom may be used to specify additional properties of the
averages (\ref{defaver:MG}), or (\ref{defaver:MG:prop}), when necessary.

Choosing different proper coordinate systems $\phi ^{i}$ will give different
average fields (\ref{defaver:MG:prop}) of a given microscopic tensor field $
p_{\beta }^{\alpha }(x)$. In general, the averages $\left\langle
p\right\rangle _{j}^{i}(\phi )$ and $\overline{p}_{j}^{i}(\widetilde{\phi })$
calculated in the proper coordinate systems $\phi ^{i}$ and $\widetilde{\phi
}^{i}$ are not related by a tensorial law under the transformation $
\widetilde{\phi }^{i}=\widetilde{\phi }^{i}(\phi ^{j})$, nor are the
operators $\mathcal{W}_{j}^{i}(\phi ^{\prime },\phi )$ and $\mathcal{W}
_{j}^{i}(\widetilde{\phi }^{\prime },\widetilde{\phi })$ related by a
tensorial transformation with each of them being $\mathbf{\delta }_{j}^{i}$
(\ref{W:delta}) in its own proper coordinates system. It should be noted here
that averages (\ref{defaver:MG}), and (\ref{defaver:MG:prop}), are obviously
tensorial with respect to coordinate transformations (as follows directly
from its definition). The reason for the \textquotedblleft
non-tensorial\textquotedblright\ properties between the proper coordinates
is due to the structure (\ref{W}), or (\ref{W:coor}), of the bilocal
operator $\mathcal{W}_{\beta }^{\alpha ^{\prime }}(x^{\prime },x)$ itself,
which involves a functional freedom in changing the functions $f_{i}^{\alpha
}(x)$, or $\phi ^{i}(x)$. This \textquotedblleft
non-tensorial\textquotedblright\ property is very natural, indeed, for it
states the exceptional character of the proper coordinate systems for
obtaining the simplest and most transparent form of the averages and the
averaging and coordination operators. It closely resembles, on the other
hand, the definition of averages in macroscopic electrodynamics as it was
noted above, and the exceptional character of Cartesian coordinates used in
that averaging procedure. The class of the proper coordinate systems on an
arbitrary differentiable metric manifold is a natural counterpart of the
Cartesian coordinate system on a Minkowski manifold. The property they share
in common is that both are volume-preserving.

There is, however, a special subclass within the class of volume-preserving
coordinate transformations described in the Proposition about fhe freedom of
the proper coordinates which keeps the bilocal operator (\ref{W:delta}) and
the averages (\ref{defaver:MG:prop}) covariant \cite{Mars-Zala:1997}.

\begin{proposition}[The covariance of the averages in the proper coordinates.
]
The class of transformations $\phi ^{i}\rightarrow \tilde{\phi}^{i}$ which
keeps the bivector $\mathcal{W}_{\beta }^{\alpha ^{\prime }}$ and the
averages (\ref{defaver:MG}) covariant within the class of proper system of
coordinates is
\begin{equation}
\tilde{\phi}^{i}=\Lambda _{j}^{i}\phi ^{j}+a^{i}  \label{Lorentz}
\end{equation}
where $\Lambda _{j}^{i}$ and $a^{i}$ are constant.
\end{proposition}

Due to this Proposition, if the manifold $(\mathcal{M}$, $g_{\alpha \beta })$
is chosen to be a (pseudo)-Riemannian space-time, the averages (\ref
{defaver:MG:prop}) defined in proper coordinates are Lorentz tensors exactly
like the averages in Minkowski space-time (see Section 2).

\section{The Parallel Transportation Bilocal Operator}

\label{*ptbo}

Let us consider a pseudo-Riemannian space-time manifold $(\mathcal{M}$, $
g_{\alpha \beta })$ of general relativity and the bilocal parallel
transportation bivector $g_{\beta }^{\alpha ^{\prime }}(x^{\prime },x)$ \cite
{Syng:1960},
\begin{equation}
g_{\beta }^{\alpha ^{\prime }}(x^{\prime },x)=t_{i}^{\alpha ^{\prime
}}(x^{\prime })t_{\beta }^{-1i}(x),  \label{parallel}
\end{equation}
which is defined for all pairs of points $(x^{\prime },x)$ on a unique
geodesic by means of the parallel-transported tetrad $t_{i}^{\alpha }(x)$.
Since the parallel transport is the fundamental operation of the
pseudo-Riemannian space-time for comparison of tensors in different points
it is important to understand if one can use this bilocal operator as the
averaging operator, $\mathcal{A}_{\beta }^{\alpha ^{\prime }}(x^{\prime
},x)=g_{\beta }^{\alpha ^{\prime }}(x^{\prime },x)$, satisfying the
properties (\ref{coincidence}) and (\ref{idempotency}) to define the
space-time average value (\ref{defaver:MG}). Upon adopting the condition
(\ref{AequalW}) to have the same bilocal operator for the space-time
averaging operator $\mathcal{A}_{\beta }^{\alpha ^{\prime }}(x^{\prime },x)$
and the coordination bivector $\mathcal{W}_{\beta }^{\alpha ^{\prime }}$,
\begin{equation}
\mathcal{A}_{\beta }^{\alpha ^{\prime }}=\mathcal{W}_{\beta }^{\alpha
^{\prime }}=g_{\beta }^{\alpha ^{\prime }},  \label{a=w=g}
\end{equation}
one must check if the differential conditions (\ref{diffW:1}) and (\ref
{diffW:2}) are satisfied for (\ref{a=w=g})\ to ensure the locality of the
space-time averages (\ref{averlocal}) and the simple formula for commutation
for operations of the averaging and the partial derivative (\ref{comm}).

The space-time averaging procedure of the macroscopic gravity based on the
Lie-dragging of tensors by means of the space-time averaging operator $
\mathcal{A}_{\beta }^{\alpha ^{\prime }}(x^{\prime },x)$ and the
coordination bivector $\mathcal{W}_{\beta }^{\alpha ^{\prime }}\ $(\ref
{coincidence}), (\ref{idempotency}) and (\ref{AequalW}) has been shown \cite
{Zala:1997}, \cite{Zala:1992}, \cite{Zala:1993}, \cite{Mars-Zala:1997}, see
Sections \ref{*sstamg}-\ref{*psc}, to satisfy the properties (\ref{diffW:2}
), (\ref{W})-(\ref{C:const}). Then the space-time averages are local
single-valued functions of the supporting point (\ref{averlocal}) and the
commutation formula (\ref{comm}) holds. In case of the definition of the
space-time average (\ref{defaver:MG}) by means of averaging operator $
\mathcal{A}_{\beta }^{\alpha ^{\prime }}(x^{\prime },x)=g_{\beta }^{\alpha
^{\prime }}(x^{\prime },x)$ (\ref{a=w=g}) the properties of the averages may
be affected by the special structure of the operators $\mathcal{A}_{\beta
}^{\alpha ^{\prime }}=g_{\beta }^{\alpha ^{\prime }}$ and $\mathcal{W}
_{\beta }^{\alpha ^{\prime }}=g_{\beta }^{\alpha ^{\prime }}$ which can be
more restrictive as compared with the space-time averaging and the
coordination bivectors based on the concept of Lie-dragging.

The space-time average value (\ref{defaver:MG}) using the parallel
transportation bivector averaging operator $g_{\beta }^{\alpha ^{\prime
}}(x^{\prime },x)$ (\ref{parallel}) can written as\footnote{
This averaging procedure is close in its construction and geometric meaning
to the the Brill-Hartle averaging scheme \cite{MTW:1973}, \cite{Isa1:1968},
\cite{Isa2:1968} for high-frequency approximation in general relativity.
However, the Brill-Hartle procedure is designed for the space-time averaging
of the weak gravitational fields with respect to the backgorund space-time
metric.}
\begin{equation}
\left\langle p\right\rangle _{\beta }^{\alpha }(x)=\frac{1}{V_{\Sigma }}
t_{i}^{\alpha }(x)t_{\beta }^{-1j}(x)\int_{\Sigma }p_{j}^{i}(x^{\prime })
\sqrt{-g^{\prime }}d^{n}x^{\prime }\equiv t_{i}^{\alpha }(x)t_{\beta
}^{-1j}(x)\langle p_{j}^{i}\rangle (x),  \label{aver-parallel}
\end{equation}
with the bilocal extension $\mathbf{p}_{\beta }^{\alpha }(x,x^{\prime })$
(\ref{bilocext}) in the integrand of (\ref{defaver:MG}) defined now as
\begin{equation}
\mathbf{p}_{\beta }^{\alpha }(x,x^{\prime })=g_{\mu ^{\prime }}^{\alpha
}(x,x^{\prime })p_{\nu ^{\prime }}^{\mu ^{\prime }}(x^{\prime })g_{\beta
}^{\nu ^{\prime }}(x^{\prime },x)=t_{i}^{\alpha }(x)t_{\beta
}^{-1j}(x)p_{j}^{i}(x^{\prime }),  \label{tetrad-componets}
\end{equation}
where $p_{j}^{i}(x^{\prime })=t_{\mu ^{\prime }}^{-1i}(x^{\prime })p_{\nu
^{\prime }}^{\mu ^{\prime }}(x^{\prime })t_{j}^{\nu ^{\prime }}(x^{\prime })$
are the tetrad components of the tensor $p_{\beta }^{\alpha }(x)$. By
choosing a tetrad $t_{i}^{\alpha }(x)$ at a supporting point $x\in \Sigma $
to which the average value will be prescribed, the tetrad is
parallel-transported at each other points $x^{\prime }\in \Sigma $ along a
geodesic connecting $x$ and $x^{\prime }$,
\begin{equation}
t_{i}^{\alpha }(x)_{;\beta }u^{\beta }=0,  \label{tetrad-parallel}
\end{equation}
where $u^{\alpha }$ is the tangent vector to a geodesic and semicolon is the
covariant derivative with respect to the Levi-Civita connection
corresponding the Riemannian metric tensor $g_{\alpha \beta }(x)$, to
construct a field of the parallel-transported tetrad in $\Sigma $. The
space-time average (\ref{aver-parallel}) is calculated by means of the
parallel transportation of the tensor $p_{\beta }^{\alpha }(x)$ along the
geodesics from each point of $x^{\prime }\in \Sigma $ to the supporting
point $x\in \Sigma $. It is easy to prove that the averaged value is
tetrad-independent.

\begin{proposition}[The tetrad-independence of the space-time averages]
The space-time average value $\left\langle p\right\rangle _{\beta }^{\alpha
}(x)$ (\ref{aver-parallel}) is invariant under the Lorentz group.
\end{proposition}

The proof is straightforward by making a Lorentz transformation $\widetilde{t
}_{i}^{\alpha }(x)=\Lambda _{i}^{j}t_{j}^{\alpha }(x)$ and observing that
the transformation matrix $\Lambda _{i}^{j}$ is constant along a geodesic
due to (\ref{tetrad-parallel}), $\Lambda _{i,\beta }^{j}u^{\beta }=0$, which
means that $\widetilde{t}_{i}^{\alpha }(x)\widetilde{t}_{\beta
}^{-1j}(x)\langle \widetilde{p}_{j}^{i}\rangle (x)=t_{i}^{\alpha
}(x)t_{\beta }^{-1j}(x)\langle p_{j}^{i}\rangle (x)$.

The space-time averaging operator $g_{\beta }^{\alpha ^{\prime }}(x^{\prime
},x)$ has the properties (\ref{coincidence}) and (\ref{idempotency}) which
follow from its definition (\ref{parallel}).

\begin{proposition}
The space-time averaging operator $\mathcal{A}_{\beta }^{\alpha ^{\prime
}}(x^{\prime },x)$ satisfies the Properties (\ref{coincidence}) and (\ref
{idempotency}): the coincidence limit of $g_{\beta ^{\prime }}^{\alpha }$ is
\begin{equation}
\lim_{x^{\prime }\rightarrow x}g_{\beta ^{\prime }}^{\alpha }(x,x^{\prime
})=\delta _{\beta }^{\alpha },  \label{coincidence-g}
\end{equation}
and it is idempotent
\begin{equation}
g_{\beta ^{\prime }}^{\alpha }(x,x^{\prime })g_{\gamma ^{\prime \prime
}}^{\beta ^{\prime }}(x^{\prime },x^{\prime \prime })=g_{\gamma ^{\prime
\prime }}^{\alpha }(x,x^{\prime \prime }).  \label{idempotency-g}
\end{equation}
\end{proposition}

Therefore the Theorem about the factorization of the averaging operator, see
Section \ref{*abao}, is also satisfied. The averaging scheme is covariant
and linear (\ref{linear}).

There is the following important lemma.

\begin{lemma}[The space-time averages of covariantly-constant tensors]
For two\linebreak\ tensor fields $p_{\beta }^{\alpha }(x)$ and $q_{\beta
}^{\alpha }(x)$ defined on $\mathcal{U}\subset \mathcal{M}$ the space-time
average (\ref{aver-parallel}) of their product $\left\langle p_{\beta
}^{\alpha }q_{\delta }^{\gamma }\right\rangle (x)$ has the form
\begin{equation}
\left\langle p_{\beta }^{\alpha }q_{\delta }^{\gamma }\right\rangle
(x)=\left\langle p\right\rangle _{\beta }^{\alpha }(x)q_{\delta }^{\gamma
}(x)  \label{covar-const}
\end{equation}
if and only if the tensor field $q_{\beta }^{\alpha }(x)$ is covariantly
constant on $\mathcal{U}\subset \mathcal{M}$,
\begin{equation}
q_{\beta ;\gamma }^{\alpha }=0.  \label{covar-const2}
\end{equation}
\end{lemma}

The proof of sufficiency is made by observation that the tetrad component $
q_{j}^{i}(x)$ in the average value are constant along the geodesics, $
q_{j,\beta }^{i}u^{\beta }=0$. The necessity follows from analysis of
(\ref{covar-const2}) with taking into account that the averaging region $\Sigma $
is arbitrary.

An important consequence of the Lemma is that the space-time average of a
covariant constant tensor $q_{\beta }^{\alpha }(x)$ (\ref{covar-const2}) is
equal to the tensor itself,
\begin{equation}
\left\langle q\right\rangle _{\beta }^{\alpha }(x)=q_{\beta }^{\alpha }(x).
\label{covar-const3}
\end{equation}

\section{The Uniqueness of the Space-time Average Value}

\label{*ustav}

In definition of the space-time average value (\ref{aver-parallel}) by the
parallel transportation bivector averaging operator $g_{\beta }^{\alpha
^{\prime }}(x^{\prime },x)$ (\ref{parallel}) it has been assumed that the
supporting point $x\in \Sigma $ is connected by only one geodesic with each
point $x^{\prime }\in \Sigma $ of the averaging region $\Sigma \subset
\mathcal{M}$. However, in the (pseudo)-Riemannian manifolds some geodesics
starting at a point may intersect in some other points of the manifold. Such
points are called conjugate \cite{Hawk-Elli:1973}. Such situations usually
occur in the presence of matter, since the matter distributions satisfying
the energy conditions due to the Einstein equations tend to focus the
timelike and null geodesics \cite{Hawk-Elli:1973}. Possible occurrence of
the points $x^{\prime }\in \Sigma $ conjugate to the supporting point $x$ of
an averaging region $\Sigma \subset \mathcal{M}$\ raises the question about
the uniqueness of space-time average values (\ref{aver-parallel}). It should
be pointed out here that according to the Whitehead theorem \cite{Whit:1932}
for the $C^{k+2}$ (pseudo)-Riemannian manifold $(\mathcal{M}$, $g_{\alpha
\beta })$, $k\geqslant 1$, each point $p$ has a normal convex neighborhood $
N_{p}$ every pair of points of which is connected by only one geodesic
belonging to the neighborhood $N_{p}$. However, a choice $\Sigma =N_{p}$
would significantly restrict the applicability of the space-time average
procedure (\ref{aver-parallel}) in the situation of interest in physically
realistic settings.

The following theorem regarding the uniqueness of the space-time averages
(\ref{aver-parallel}) takes place \cite{Arif-Zala:1989}.

\begin{theorem}[The uniqueness of the space-time averages]
Let a $C^{k}$ tensor field\linebreak\ $p_{\beta }^{\alpha }(x)$, $k\geqslant
1$, $x\in \mathcal{M}$, is given on a region $\mathcal{U}\subset \mathcal{M}$
of a (pseudo)-Riemannian manifold $(\mathcal{M}$, $g_{\alpha \beta })$. Then
for an arbitrary region $\Sigma \subset \mathcal{U}$\ the space-time average
value $\left\langle p\right\rangle _{\beta }^{\alpha }(x)$ (\ref
{aver-parallel}) at the supporting point $x\in \Sigma $ is uniquely
determined.
\end{theorem}

Proof is as follows. If there are no conjugate points $x^{\prime }$ in the
region $\Sigma $ for the supporting point $x\in \Sigma $, the average value
is uniquely determined since there is only one geodesic connecting $x$ with
each point $x^{\prime }\in \Sigma $. Let us assume now that there are some
points $x^{\prime }\in \Sigma $ conjugate to $x\in \Sigma $. It is known
that the points $x^{\prime }\in \Sigma $ of intersection of geodesics can
only form a subspace of the dimension $\,0\leqslant m\leqslant n-1$, where $
n=\dim \mathcal{M}$ \cite{DeWi:1967}. Consider a geodesic between the
supporting point $x$ and a point $x^{\prime }$ and introduce the world
function of Synge \cite{Syng:1960}
\begin{equation}
\sigma (x,x^{\prime })=\frac{\tau ^{\prime }-\tau }{2}\int\limits_{x}^{x^{
\prime }}g_{\alpha \beta }u^{\alpha }u^{\beta }d\tau ,\quad \sigma
(x,x^{\prime })=\sigma (x^{\prime },x),  \label{synge-func}
\end{equation}
where $\tau ^{\prime }$ and $\tau $ are values of the geodesic parameter at
points $x^{\prime }$ and $x$, correspondingly. The biscalar of the world
function $\sigma (x,x^{\prime })$ satisfy the fundamental equations
following from the Hamilton-Jacobi equations of the action $S=2\sigma
(x,x^{\prime })/(\tau ^{\prime }-\tau )$ for each point,
\begin{equation}
\frac{1}{2}\sigma _{\alpha }\sigma ^{\alpha }=\sigma ,\quad \frac{1}{2}
\sigma _{\alpha ^{\prime }}\sigma ^{\alpha ^{\prime }}=\sigma ,
\label{hamilton-jacobi}
\end{equation}
where $\sigma _{\alpha }=\partial \sigma /\partial x^{\alpha }$ and $\sigma
_{\alpha ^{\prime }}=\partial \sigma /\partial x^{\alpha ^{\prime }}$ are
the vectors tangent to the geodesic at point $x$ and $x^{\prime }$ having
the length equal to the parametric distance between $x^{\prime }$ and $x$ in
accordance with (\ref{hamilton-jacobi}). Let us consider now a geodesic from
$x$ to a point $x^{\prime \prime }$ close to $x^{\prime }$, $\delta
x^{\prime }=x^{\prime \prime }-x^{\prime }$. The change of the coordinates $
\delta x^{\prime }$ due to a variation of $\sigma _{\alpha }$ is determined
by $\delta \sigma _{\alpha }=\sigma _{\alpha ,\mu ^{\prime }}\delta x^{\mu
^{\prime }}$. This equation is invertible, $\delta x^{\mu ^{\prime }}=\sigma
^{-1\mu ^{\prime }\alpha }\delta \sigma _{\alpha }$\ , due to $\det (\sigma
_{\alpha ,\mu ^{\prime }})\neq 0$, since $\sigma _{\alpha ,\mu ^{\prime
}}\rightarrow g_{\alpha \mu }$ as $x^{\prime }\rightarrow x$ \cite{Syng:1960}
. If $x^{\prime }$ is a conjugate point to $x$ then $\delta x^{\mu ^{\prime
}}=0$ and the following system of the linear homogeneous algebraic equations
$\sigma ^{-1\mu ^{\prime }\alpha }\delta \sigma _{\alpha }=0$ has a
nontrivial solution if
\begin{equation}
\det (\sigma ^{-1\mu ^{\prime }\alpha })=0.  \label{conjugate-system}
\end{equation}
Therefore, with a fixed supporting point $x\in \Sigma $ the points $
x^{\prime }\in \Sigma $ where the geodesics are intersecting occupy at most
a hypersurface (\ref{conjugate-system}). The proof is essentially local
since the variation equation $\delta \sigma _{\alpha }=\sigma _{\alpha ,\mu
^{\prime }}\delta x^{\mu ^{\prime }}$ is defined only in a neighborhood of $
x^{\prime }$ and $x$, but it suffices the goal of the theorem because the
averaging procedure (\ref{defaver:MG}) and (\ref{aver-parallel}) is local as
well.

It follows from theorem that if there are conjugate points in the averaging
region $\Sigma \subset \mathcal{M}$ they form a 3-dimensional subspace of
measure zero at the 4-dimensional averaging integral (\ref{aver-parallel}),
which ensures the uniqueness of the space-time averages.

Thus, the algebra of the space-time average value (\ref{aver-parallel})
using the parallel transportation averaging bivector operator $g_{\beta
}^{\alpha ^{\prime }}(x^{\prime },x)$ (\ref{parallel}) is very similar to
the algebra of the space-time averaging procedure of macroscopic gravity,
see Sections \ref{*sstamg} and \ref{*abao}, with an additional Lemma, see
Section \ref{*ptbo}, following from the properties of the averaging operator
$g_{\beta }^{\alpha ^{\prime }}(x^{\prime },x)$.

\section{The Differential Properties of the Space-time Averages}

\label{*dpata}

The differential properties of the space-time average value (\ref
{aver-parallel}) using the parallel transportation averaging bivector
operator $g_{\beta }^{\alpha ^{\prime }}(x^{\prime },x)$ are not simple as
compared with the space-time averaging procedure of macroscopic gravity, see
Sections \ref{*dpbo} and \ref{*et}.

As a consequence of the definitions (\ref{derivative}) and (\ref{shift})
with (\ref{parallel}), (\ref{a=w=g}) and (\ref{aver-parallel}) one can
obtain the following formula for the commutation of the covariant
differentiation and the space-time averaging \cite{AZS:1990a}:
\begin{equation}
\left\langle p\right\rangle _{\beta ;\lambda }^{\alpha }=\langle g_{\mu
^{\prime }}^{\alpha }p_{\nu ^{\prime };\epsilon ^{\prime }}^{\mu ^{\prime
}}g_{\beta }^{\nu ^{\prime }}g_{\lambda }^{\epsilon ^{\prime }}\rangle
+\langle \mathbf{p}_{\beta }^{\alpha }g_{\lambda ;\epsilon ^{\prime
}}^{\epsilon ^{\prime }}\rangle -\left\langle p\right\rangle _{\beta
}^{\alpha }\langle g_{\lambda ;\epsilon ^{\prime }}^{\epsilon ^{\prime
}}\rangle -\langle \mathcal{G}^{\alpha }{}_{\sigma \lambda }\mathbf{p}
_{\beta }^{\sigma }\rangle +\langle \mathbf{p}_{\sigma }^{\alpha }\mathcal{G}
^{\sigma }{}_{\beta \lambda }\rangle .  \label{covar-averaging}
\end{equation}
Here $\mathcal{G}^{\alpha }{}_{\sigma \lambda }$ are the structural
functions,
\begin{equation}
\mathcal{G}^{\alpha }{}_{\beta \gamma }=g_{\epsilon ^{\prime }}^{\alpha }
\mathcal{G}^{\epsilon ^{\prime }}{}_{\beta \gamma }=g_{\epsilon ^{\prime
}}^{\alpha }(g_{\beta ;\gamma }^{\epsilon ^{\prime }}+g_{\beta ;\sigma
^{\prime }}^{\epsilon ^{\prime }}g_{\gamma }^{\sigma ^{\prime }}),
\label{structure-parallel}
\end{equation}
and $g_{\lambda ;\epsilon ^{\prime }}^{\epsilon ^{\prime }}$ is the
divergence of the parallel transportation bivector, $g_{\lambda ;\epsilon
^{\prime }}^{\epsilon ^{\prime }}=g_{\lambda ,\epsilon ^{\prime }}^{\epsilon
^{\prime }}+\Gamma _{\epsilon ^{\prime }}g_{\lambda }^{\epsilon ^{\prime }}$
, $\Gamma _{\epsilon ^{\prime }}=\Gamma _{\epsilon ^{\prime }\alpha ^{\prime
}}^{\alpha ^{\prime }}=(\ln \sqrt{-g})_{,\epsilon ^{\prime }}$ where $\Gamma
_{\beta \gamma }^{\alpha }$ are the Levi-Civita connection coefficients $
\Gamma _{\beta \gamma }^{\alpha }$. Despite its complicated form the
commutation formula (\ref{covar-averaging}) has a very transparent meaning:
the first term in the right-hand side is the average value of the derivative
of the covariant derivative $p_{\beta ;\gamma }^{\alpha }$, the last two
terms are due to the nontrivial structural functions for the averaging
operator $g_{\beta }^{\alpha ^{\prime }}$, while the second and third terms
are present because of the parallel transportation bivector is not defined
in general along a volume-preserving congruence of geodesics. The formula
for the commutation of the partial differentiation and the averaging can be
easily written on the basis of the commutation formula (\ref{covar-averaging}
) by changing the covariant derivative at the supporting point $x$ by the
partial derivative.

The general form of the structural functions is given by
\begin{equation}
\mathcal{G}^{\alpha ^{\prime }}{}_{\beta \gamma }=\int\limits_{\tau
_{0}}^{\tau ^{\prime }}d\tau g_{\beta }^{\mu ^{\prime \prime }}r^{\epsilon
^{\prime \prime }}{}_{\mu ^{\prime \prime }\nu ^{\prime \prime }\rho
^{\prime \prime }}g_{\epsilon ^{\prime \prime }}^{\alpha ^{\prime }}u^{\nu
^{\prime \prime }}\left( \mathcal{X}^{\rho ^{\prime \prime }}{}_{\gamma }+
\mathcal{Y}^{\rho ^{\prime \prime }}{}_{\sigma ^{\prime }}g_{\gamma
}^{\sigma ^{\prime }}\right)   \label{structure-parallel2}
\end{equation}
where $r^{\alpha }{}_{\beta \gamma \delta }(x)$ is the Riemann curvature
tensor \cite{Scho:1954} and $\mathcal{X}^{\rho ^{\prime \prime }}{}_{\gamma }
$ and $\mathcal{Y}^{\rho ^{\prime \prime }}{}_{\sigma ^{\prime }}$ are the
trilocal functions of solutions to the Jacobi equation. They do not vanish
in the pseudo-Riemannian space-time manifold and can be expressed in terms
of the Riemann curvature tensor . The operations of the covariant (partial)
differentiation and the space-time averaging do not commute even if one
requires the coordination parallel transportation bivector $g_{\beta
}^{\alpha ^{\prime }}$ to be volume-preserving,
\begin{equation}
g_{\lambda ;\epsilon }^{\epsilon }=0,\quad g_{\lambda ;\epsilon ^{\prime
}}^{\epsilon ^{\prime }}=0,  \label{volume-preserving}
\end{equation}
due to the nonvanishing of the terms involving the structural functions $
\mathcal{G}_{\sigma \lambda }^{\alpha }$ in (\ref{covar-averaging}). To
guarantee the locality (\ref{averlocal}) of the space-time averages, see
Section \ref{*dpbo}, one needs to satisfy the existence Theorems of Sections
\ref{*et} to require
\begin{equation}
g_{[\beta ,\gamma ]}^{\alpha ^{\prime }}+g_{[\beta ,\delta ^{\prime
}}^{\alpha ^{\prime }}g_{\gamma ]}^{\delta ^{\prime }}=0,
\label{locality-parallel}
\end{equation}
for the parallel transportation bivector $g_{\beta }^{\alpha ^{\prime }}$
(\ref{parallel}) such that the parallel-transported tetrad $t_{i}^{\alpha
}(x)\partial _{\alpha }=\mathbf{t}_{i}$ has the commutation relations
\begin{equation}
\left[ \mathbf{t}_{i},\mathbf{t}_{j}\right] =C_{ij}^{k}\mathbf{t}_{k}
\label{commutator-parallel}
\end{equation}
with the constant structure functions (anholonomicity coefficients) $
C_{ij}^{k}$,
\begin{equation}
C_{ij}^{k}=\mathrm{const}.  \label{unholonomicity-parallel}
\end{equation}

If the conditions (\ref{locality-parallel})-(\ref{unholonomicity-parallel})
are satisfied, the space-time average (\ref{aver-parallel}) is a
single-valued local function of the supporting point $x$. With the operator
being volume-preserving (\ref{volume-preserving}), the average tensor field
does not depend explicitly on the value of the region's volume $V_{\Sigma }$,
and $V_{\Sigma }$ itself is a free parameter of the averages.

The requirement for the averaging operator $g_{\beta }^{\alpha ^{\prime
}}(x^{\prime },x)$ to be volume-preserving (\ref{volume-preserving}) leads
\cite{Zala:unpub} to the class of D'Atri spaces (see, for example, \cite
{Tric-Vanh:1983} for definitions and relevant results), which is a special
class of the (pseudo)-Riemannian spaces with particular restrictions of
their curvature. The locality conditions (\ref{locality-parallel})-(\ref
{unholonomicity-parallel}) put additional conditions on the structure of
geodesics and the curvature tensor.

In should be pointed here that the volume-preserving bases $f_{i}^{\alpha
}(x)\partial _{\alpha }=\mathbf{f}_{i}$ (\ref{diffW:1}), (\ref{averlocal})
and (\ref{diffW:2}) with the constant anholonomicity coefficients defining
the averaging and coordination operators $\mathcal{A}_{\beta }^{\alpha
^{\prime }}(x^{\prime },x)$ and $\mathcal{W}_{\beta }^{\alpha ^{\prime
}}(x^{\prime },x)$ (\ref{AequalW}) for the space-time averaging scheme (\ref
{defaver:MG}) adopted in macroscopic gravity, exist always, at least,
locally, on any (pseudo)-Riemannian space-times without any restrictions on
the curvature\footnote{
It follows from the well-known fact that on any (pseudo)-Riemannian space
there always exists a coordinate system in which the connection coefficients
$\gamma _{\beta \delta }^{\alpha }$ have $\gamma _{\beta \alpha }^{\alpha }=0
$, or, what is the same, $\det (g_{\alpha \beta })=\mathrm{const}$.}. This
averaging scheme, therefore, in addition to the possibility to define
averaged fields (\ref{defaver:MG}) with reasonable algebraic and
differential properties, is applicable locally, as discussed above, on any
space-time manifold. This is an essential advantage of the scheme, which
allows one to consider it, as well as the results of its application for the
space-time averaging of the pseudo-Riemannian geometry and general
relativity, as being generic from both geometrical and physical points of
view\footnote{
This analysis gives an indication, in our opinion, that using such bivectors
as, for instance, the parallel transportation bivector, for the averaging of
general relativity implies another set of basic assumptions about the nature
of averaged gravity and the character of space-time measurements \cite
{Zala:unpub}. It also may be related to a quantum regime of gravitation as a
physical setting where such averaging operators are more adequate (see, for
example, \cite{DeWi:1962}). For formulation of a classical macroscopic
theory of gravity, the proposed space-time averaging procedure based on a
Lie-dragging model of space-time measurements is relevant and it is a
simplest generalization of the flat space-time procedure adopted in
hydrodynamics and macroscopic electrodynamics.}.

The space-time averaging scheme being considered here is essentially of
local character in the sense that the average values are defined by  (\ref
{defaver:MG}) and (\ref{aver-parallel}) over local regions $\Sigma $ of a
microscopic manifold $\mathcal{M}$, and thereby the average fields are
defined locally on $\mathcal{U}$ with the topological and differentiable
structure of $\mathcal{M}$ remaining unchanged. Such local character of the
macroscopic picture is dictated, first of all, from the physical point of
view, by our experience and observations which show that physical quantities
are represented by local functions determined by means of measurements which
are themselves fundamentally of local character (i.e., a measurement of a
physical quantity is carried out always during a finite time period over a
finite space region, to be small compared with the characteristic extension
of the system under interest and its time of existence). Thus, from the
mathematical point of view, to describe such objects adequately, it is
sufficient to formulate a calculus of the averages on a differentiable
manifold. If such a calculus is formulated, a definition of an average field
globally can be done in the same way as one constructs a global field on a
manifold if it possesses a nontrivial topology.

\section{The Non-Riemannian Geometry of the Averaged Space-time}

Due to the properties of the space-time average value (\ref{aver-parallel})
using the parallel transportation averaging bivector operator $g_{\beta
}^{\alpha ^{\prime }}(x^{\prime },x)$, see Sections \ref{*ptbo}-\ref{*dpata}
there are two fundamental theorems following from the Lemma of Section \ref
{*ptbo}.

\begin{theorem}[The averaged metric tensor]
The space-time average (\ref{aver-parallel}) of the metric tensor $g_{\alpha
\beta }(x)$ of the pseudo-Riemannian geometry of general relativity is equal
to the metric tensor itself,
\begin{equation}
g_{\alpha \beta ;\gamma }=0\quad \rightarrow \quad \left\langle g_{\alpha
\beta }\right\rangle (x)=g_{\alpha \beta }(x).  \label{aver-metric}
\end{equation}
\end{theorem}

\begin{theorem}[Averaging out the symmetric space-times]
The space-time average\linebreak\ (\ref{aver-parallel}) of the symmetric
space-times of the pseudo-Riemannian geometry of general relativity leave
them unchanged ,
\begin{equation}
r^{\alpha }{}_{\beta \gamma \delta ;\epsilon }=0\quad \rightarrow \quad
\left\langle r^{\alpha }{}_{\beta \gamma \delta }\right\rangle (x)=r^{\alpha
}{}_{\beta \gamma \delta }(x).  \label{aver-symmetric}
\end{equation}
\end{theorem}

The theorem establishing the algebraic properties of the averaged Riemann
curvature tensor, $\left\langle r^{\alpha }{}_{\beta \gamma \delta
}\right\rangle (x)=R^{\alpha }{}_{\beta \gamma \delta }(x)$, from the
algebraic properties of the Riemann curvature tensor of the
pseudo-Riemannian geometry \cite{Scho:1954} of general relativity
\begin{equation}
r^{\alpha }{}_{\beta (\gamma \delta )}=0,\quad r^{\alpha }{}_{[\beta
\gamma \delta ]}=0,\quad g_{\alpha \epsilon }r^{\epsilon }{}_{\beta \gamma
\delta }+g_{\beta \epsilon }r^{\epsilon }{}_{\alpha \gamma \delta
}=r_{(\alpha \beta )\gamma \delta }=0,  \label{curvature-identities}
\end{equation}
where the round and square brackets stand for the index symmetrization and
antisymmetrization, correspondingly, can be also proved in the basis of the
algebra of the space-time averages.

\begin{theorem}[Algebraic properties of the averaged curvature tensor]
There are\linebreak\ the following algebraic identities on the averaged
Riemann curvature tensor, $\left\langle r^{\alpha }{}_{\beta \gamma \delta
}\right\rangle (x)=R^{\alpha }{}_{\beta \gamma \delta }(x)$,
\begin{equation}
R^{\alpha }{}_{\beta (\gamma \delta )}=0,\quad R^{\alpha }{}_{[\beta
\gamma \delta ]}=0,\quad g_{\alpha \epsilon }R^{\epsilon }{}_{\beta \gamma
\delta }+g_{\beta \epsilon }R^{\epsilon }{}_{\alpha \gamma \delta }=0.
\label{curvature-aver-identities}
\end{equation}
\end{theorem}

\begin{corollary}[The averaged Ricci tensor is symmetric]
The averaged Ricci tensor $R_{\alpha \beta }=R^{\epsilon }{}_{\alpha \beta
\epsilon }$ is symmetric,
\begin{equation}
R_{\alpha \beta }=R_{\beta \alpha }.  \label{ricci-symmetric}
\end{equation}
\end{corollary}

The obtained results allow one to establish the formal structure of the
space-time averging out of the Einstein equations
\begin{equation}
r_{\alpha \beta }-\frac{1}{2}g_{\alpha \beta }g^{\mu \nu }r_{\mu \nu }=
- 8\pi kt_{\alpha \beta }^{\mathrm{(micro)}}  \label{einstein-micro}
\end{equation}
where $r_{\alpha \beta }=r^{\epsilon }{}_{\alpha \beta \epsilon }$ is the
Ricci tensor and $t_{\alpha \beta }^{\mathrm{(micro)}}$ is the
energy-momentum tensor of a microscopic point-like distribution.

\begin{theorem}[The averaged Einstein equations]
The space-time averaging out of the Einstein equations (\ref{einstein-micro}
) by means of (\ref{aver-parallel}), (\ref{covar-const}) and (\ref
{aver-metric}) gives
\begin{equation}
R_{\alpha \beta }-\frac{1}{2}g_{\alpha \beta }g^{\mu \nu }R_{\mu \nu }=
- 8\pi k\left\langle t_{\alpha \beta }^{\mathrm{(micro)}}\right\rangle .
\label{einstein-aver}
\end{equation}
\end{theorem}

There is a corollary for the space-time averaging out of the vacuum
space-times,
\begin{equation}
r_{\alpha \beta }=0.  \label{vacuum}
\end{equation}

\begin{corollary}[Averaged vacuum Einstein equations]
The space-time averaging out of the vacuum Einstein equations (\ref{vacuum})
by means of (\ref{aver-parallel}) gives the macrovacuum equations
\begin{equation}
R_{\alpha \beta }=0.  \label{einstein-vacuum}
\end{equation}
\end{corollary}

Now one can formulate the theorem regarding the space-time averaging out of
the vacuum space-times (\ref{vacuum}) which states that the geometry of the
averaged space-time is non-Riemannian in general \cite{AZS:1990a}.

\begin{theorem}[The non-Riemannian averaged vacuum space-times]
The macro- \linebreak vacuum space-time (\ref{einstein-vacuum}) with the
Riemann curvature tensor $R^{\alpha }{}_{\beta \gamma \delta }$ (\ref
{curvature-aver-identities}) of the Riemannian metric tensor $G_{\alpha
\beta }$ is pseudo-Riemannian if and only if $R^{\alpha }{}_{\beta \gamma
\delta }=0$ or $R^{\alpha }{}_{\beta \gamma \delta }=r^{\alpha }{}_{\beta
\gamma \delta }$.
\end{theorem}

The theorem is proved by the simultaneous analysis of the algebraic
properties (\ref{curvature-aver-identities}) of the averaged tensor $
R^{\alpha }{}_{\beta \gamma \delta }$ which is assumed by the conditions of
the Theorem to be the Riemann curvature tensor of the Riemannian metric
tensor $G_{\alpha \beta }$, and the algebraic properties of the curvature
tensor $R^{\alpha }{}_{\beta \gamma \delta }$. The identities following from
the covariant constancy of the metric tensor $G_{\alpha \beta }$,
\begin{equation}
G_{\alpha \epsilon }R^{\epsilon }{}_{\beta \gamma \delta }+G_{\beta \epsilon
}R^{\epsilon }{}_{\alpha \gamma \delta }=R_{(\alpha \beta )\gamma \delta }=0,
\label{metric-curvature}
\end{equation}
are analyzed by means of considering all possible Segre algebraic types of
the second rank symmetric tensor $g_{\alpha \beta }$ which is not equal to $
G_{\alpha \beta }$ in general, $g_{\alpha \beta }\neq G_{\alpha \beta }$.

This result is very important for understanding the structure of the
averaged gravitation, since it shows that the pseudo-Riemannian geometry of
general relativity is not preserved under the space-time averaging (\ref
{aver-parallel}) of the space-time manifolds.

\section{Conclusions}

(1) The covariant space-time averaging procedure for objects (tensors,
geometric objects, etc.) in the framework of macroscopic gravity \cite
{Zala:1997}, \cite{Zala:1992}, \cite{Zala:1993} is a natural generalization
of the space-time averaging procedure of the classical Maxwell macroscopic
electrodynamics. It gives a covariant formulation of the conditions on the
averages, which provide them with natural algebraic and analytical
properties. A wide class of averaging and coordination bilocal operators
satisfying all the properties exit locally on an arbitrary $n$-dimensional
differentiable manifold with a volume $n$-form, including metric and affine
connection manifolds and, in particular, (pseudo)-Riemannian spaces. The
class of proper coordinate systems, analogous to the Cartesian coordinate
system of the Minkowski space-time and generalizing them, gives the simplest
and most transparent form of the averages. The averaging procedure as it is
formulated allows a large functional freedom which is incorporated in an
elegant way and can be used to arrange additional specific conditions for
the averages.

(2) The covariant averaging procedure by means of using the parallel
transportation averaging bivector operator is another generalization of the
space-time averaging procedure of the classical Maxwell macroscopic
electrodynamics. This averaging scheme is much more restrictive than the
procedure of macroscopic gravity. In general, such space-time average values
do not satisfy the locality conditions, they depend on the averaging volume
explicitly and the operations of averaging and taking the covariant
(partial) derivatives do not commute. To ensure some of these properties one
needs to put strong restrictions on the Riemann curvature tensor of the
pseudo-Riemannian space-time manifolds. On the contrary, the averaging
scheme of macroscopic gravity makes it possible to define the averaged
fields on any space-time manifold such that they are local single-valued
functions, depend on the averaging volume value as a parameter and the
operations of averaging and taking the covariant (partial) derivatives do
commute. This is an essential advantage of the scheme, which allows one to
consider it, as well as the results of its application for the space-time
averaging of the pseudo-Riemannian geometry and general relativity, as being
generic from both geometrical and physical points of view.

(3) The covariant space-time averaging of macroscopic gravity has been shown
to lead to the non-Riemannian averaged space-time geometry \cite{Zala:1997},
\cite{Zala:1992}, \cite{Zala:1993}. The covariant space-time averaging
procedure using the parallel transportation averaging bivector operator also
leads to the non-Riemannian averaged space-time geometry. This result is of
very significant value since it shows that even having generically different
analytic and algebraic properties, both procedures, however, do not preserve
the pseudo-Riemannian geometry of general relativity under the space-time
averaging of space-time manifolds.

These results urge the necessity to understand the geometrical and physical
foundations of these space-time averaging schemes. The following issues are
of particular importance here:

$\bullet $ to understand the geometric meaning of these space-time averaging
procedures for the pseudo-Riemannian space-time manifolds,

$\bullet $ to make further comparison of both procedures to find out the
similarities and differences between them and the consequences of their application
for averaging out of the pseudo-Riemannian space-time manifolds,

$\bullet $ to find out and describe the classes of the pseudo-Riemannian
space-time manifolds which allow one to ensure the proper differential
properties for the space-time averages defined by the parallel
transportation averaging bivector operator,

$\bullet $ to formulate the corresponding models of the space-time
measurements based on the space-time averaging procedures of macroscopic
gravity and the parallel transportation averaging bivector operator,

$\bullet $  to understand finally the physical consequences of using both
averaging schemes for the construction of a realistic cosmological model
which is inhomogeneous on local scales of matter structures and homogeneous
and isotropic on the largest cosmological scales.


\newpage

\begin{thebibliography}{99}
\bibitem{Nova:1955} V. Novacu, \emph{Introducere \^\i n Electrodinamic\u a}
(Bucharest, Editura Academiei, 1955) (in Romanian).

\bibitem{Pano-Phil:1962} W.K.H. Panovsky and M. Phillips, \emph{Classical
Electricity and Magnetism} (Addison-Wesley, Reading, 1962).

\bibitem{Rose:1965} L. Rosenfeld, \emph{Theory of Electrons }
(Addison-Wesley, Reading, 1962).

\bibitem{Inga-Jami:1985} R.S. Ingarden and A. Jamio\l kowski, \emph{
Classical Electrodynamics} (PWN, Warszawa and Elsevier, Amsterdam, 1985).

\bibitem{deGr:1969} S.R. de Groot, \emph{The Maxwell Equations}
(North-Holland, Amsterdam, 1969).

\bibitem{Kamp:1956} J. Kampe de F\'{e}riet, in \emph{Proc. Inter. Congress
Math. Amsterdam, 1954}, Pt. 3 (North-Holland, Amsterdam, 1956), p. 237.

\bibitem{Rota:1960} G.C. Rota, \emph{Proc. Nat. Acad. Sci. USA} \textbf{46}
(1960) 863.

\bibitem{Russ:1970} G. Russakoff, \emph{Amer. J. Phys.} \textbf{38} (1970)
1188.

\bibitem{Robi:1973} F.N.H. Robinson, \emph{Macroscopic Electrodynamics}
(Pergamon Press, Oxford, 1973).

\bibitem{Jack:1975} J.D. Jackson, \emph{Classical Electrodynamics} (John
Wiley \& Sons, New York, 1975).

\bibitem{Zala:1997} R.M. Zalaletdinov, \emph{Bull. Astr. Soc. India} \textbf{
25} (1997) 401.

\bibitem{Bohr-Rose:1933} N. Bohr and L. Rosenfeld, \emph{Mat.-fys. Medd.
Dan. Vid. Selsk.} \textbf{12}, no. 8 (1933) [English translation in \emph{
Selected Papers of L\'eon Rosenfeld}, eds. R.S. Cohen and J.J. Stachel (D.
Reidel, Dordrecht, 1979) p. 357].

\bibitem{DeWi:1962} B.S. DeWitt, in \emph{Gravitation: An introduction to
current research}, ed. L. Witten (Wiley, New York, 1962), p. 266.

\bibitem{MTW:1973} C.W. Misner, K.S. Thorne and J.A. Wheeler,\emph{\
Gravitation} (Freeman, San Francisco, 1973).

\bibitem{Elli:1984} G.F.R. Ellis, in \emph{General Relativity and
Gravitation }, eds. B. Bertotti, F. de Felici and A. Pascolini (Reidel,
Dordrecht, 1984), p. 215.

\bibitem{Zala:1992} R.M. Zalaletdinov, \emph{Gen. Rel. Grav.} \textbf{24}
(1992) 1015.

\bibitem{Kras:1996} A. Krasi\'{n}ski, \emph{Inhomogeneous Cosmological Models
} (Cambridge University Press, Cambridge, 1997).

\bibitem{Zala:1993} R.M. Zalaletdinov, \emph{Gen. Rel. Grav.} \textbf{25}
(1993) 673.

\bibitem{Zala:1994} R.M. Zalaletdinov, in: \emph{Proceedings of
International Symposium on Experimental Gravitation}, eds. M. Karim and A.
Qadir (IOP, Bristol, 1994) p. A363.

\bibitem{Zala:1995} R.M. Zalaletdinov, in: \emph{Inhomogeneous Cosmological
Models}, eds. A. Molina and J.M.M. Senovilla (World Scientific, Singapore,
1995), p. 91.

\bibitem{Zala:1996} R.M. Zalaletdinov, in: \emph{Proc. of the 7th Marcel
Grossmann Meeting,} Part A, eds. R.T. Jantzen and G. Mac Keiser (World
Scientific, Singapore, 1996), p. 394.

\bibitem{Mars-Zala:1997} M. Mars and R.M. Zalaletdinov, \emph{J. Math. Phys.}
\textbf{38} (1997) 4741.

\bibitem{Schu:1980} B.F. Schutz, \emph{Geometrical Methods of Mathematical
Physics} (Cambridge University Press, Cambridge, 1980).

\bibitem{Koba-Nomi:1963} S. Kobayashi and K. Nomizu, \emph{Foundations of
Differential Geometry}, Vol. I (Interscience, New York, 1963).

\bibitem{Moni-Yagl:1971} A.S. Monin and A.M. Yaglom, \emph{Statistical Fluid
Mechanics: Mechanics of Turbulence}, Vol. 1 (Nauka, Moscow, 1965) (in
Russian) [English translation of Vol.1 revised by the authors, ed. J.L.
Lumley (MIT, Cambridge, Mass., 1971)].

\bibitem{Zala:unpub} R.M. Zalaletdinov, unpublished.

\bibitem{Lore:1916} H.A. Lorentz, \emph{The Theory of Electrons} (Teubner,
Leipzig, 1916).

\bibitem{Lesl:1973} C. Leslie, \emph{Develoments in the Theory of Turbulence}
(Claredon Press, Oxford, MA, 1973).

\bibitem{Stan:1985} M.M. Stani\v{s}i\'{c}, \emph{The Mathematical Theory of
Turbulence} (Springer-Verlag, New York, 1985).

\bibitem{AZS:1990b} L.Ya. Arifov, R.M. Zalaletdinov and A.V. Shein, \emph{
Preprint}, Institute of Nuclear Physics, Academy of Sciences of UzSSR, No.
R-12-499 (Tashkent, 1990) (in Russian).

\bibitem{AZS:1990a} L.Ya. Arifov, R.M. Zalaletdinov and A.V. Shein, \emph{
Preprint}, Institute of Nuclear Physics, Academy of Sciences of UzSSR, No.
R-12-480 (Tashkent, 1990) (in Russian).

\bibitem{deRh:1960} G. de Rham, \emph{Vari\'et\'es Diff\'erentiables},
Actualit\'es Scientifiques et Industrielles 1222 (Hermann, Paris, 1960).

\bibitem{Mose:1965} J. Moser,\emph{\ Trans. Amer. Math. Soc. }\textbf{120}
(1965) 286.

\bibitem{Grom:1986} M. Gromov, \emph{Partial Differential Relations}
(Springer-Verlag, Berlin, 1986).

\bibitem{Syng:1960} J.L. Synge, \emph{Relativity The General Theory}
(North-Holland, Amsterdam, 1960).

\bibitem{Isa1:1968} R.A. Isaacson, \emph{Phys. Rev.} \textbf{166} ({1968) }
1263 .

\bibitem{Isa2:1968} R.A. Isaacson, \emph{Phys. Rev.} \textbf{166} ({1968) }
1272 .

\bibitem{Hawk-Elli:1973} S. Hawking and G.F.R. Ellis, \emph{The Large Scale
Structure of Space-time} (Cambridge University Press, Cambridge, 1973).

\bibitem{Whit:1932} J.H.C. Whitehead, \emph{Quart. J. Math. (Oxford Ser.)}
\textbf{3} (1932) 33.

\bibitem{Arif-Zala:1989} L.Ya. Arifov and R.M. Zalaletdinov, \emph{Preprint}
, Institute of Nuclear Physics, Academy of Sciences of UzSSR, No. R-12-450
(Tashkent, 1989) (in Russian).

\bibitem{DeWi:1967} B.S. DeWitt, \emph{Phys. Rev.} \textbf{162} ({1967) }
1195.

\bibitem{Scho:1954} J.A. Schouten, \emph{Ricci-calculus} (Springer-Verlag,
Berlin, 1954).

\bibitem{Tric-Vanh:1983} F. Tricerri and L. Vanhecke, \emph{Homogeneous
Structures on Riemannian Manifolds} (Cambridge University Press, Cambridge,
1983).
\end{thebibliography}
\end{document}